\documentclass[letterpaper,12pt]{article}
\usepackage{latexsym}
\usepackage[left=1in,top=1in,right=1in,bottom=1in]{geometry}
\usepackage{amsmath}
\usepackage{color}
\usepackage{hyperref}
\usepackage{graphicx}
\graphicspath{{simulation-res/}}
\usepackage{caption}
\usepackage{subcaption}
\usepackage{amssymb}
\usepackage{float}
\usepackage{longtable}
\usepackage{amsmath}
\usepackage{algorithm} 
\usepackage{algpseudocode}
\usepackage{algorithmicx}
\DeclareMathOperator{\E}{\mathbb{E}}

\DeclareMathOperator*{\argmin}{arg\,min}
\DeclareMathOperator*{\expit}{expit}
\DeclareMathOperator*{\logit}{logit}
\newcommand{\1}{{\bf 1}}
\newcommand{\Bernoulli}{{\rm Bernoulli}}
\newcommand{\cvBias}{{\rm cvBias}}
\newcommand{\cvRSS}{{\rm cvRSS}}
\newcommand{\cvVar}{{\rm cvVar}}
\newcommand{\Val}{{\rm Val}}
\newcommand{\xL}{{\cal L}}

\definecolor{darkblue}{rgb}{0,0.4,0.9}
\definecolor{gray10}{rgb}{0.1,0.1,0.1}
\definecolor{gray20}{rgb}{0.2,0.2,0.2}
\definecolor{gray30}{rgb}{0.3,0.3,0.3}
\definecolor{gray40}{rgb}{0.4,0.4,0.4}
\definecolor{gray60}{rgb}{0.6,0.6,0.6}
\definecolor{gray80}{rgb}{0.8,0.8,0.8}
\definecolor{gray90}{rgb}{0.9,0.9,.9}
\definecolor{gray95}{rgb}{0.95,0.95,.95}
\definecolor{gray96}{rgb}{0.96,0.96,.96}
\definecolor{lgreen} {RGB}{180,210,100}
\definecolor{dblue}  {RGB}{20,66,129}
\definecolor{ddblue} {RGB}{11,36,69}
\definecolor{lred}   {RGB}{220,0,0}
\definecolor{nred}   {RGB}{224,0,0}
\definecolor{norange}{RGB}{230,120,20}
\definecolor{nyellow}{RGB}{255,221,0}
\definecolor{ngreen} {RGB}{98,158,31}
\definecolor{dgreen} {RGB}{78,138,21}
\definecolor{nblue}  {RGB}{28,130,185}
\definecolor{jblue}  {RGB}{20,50,100}
\definecolor{nnyellow}{RGB}{235,200,0}
\definecolor{purple}{RGB}{150, 0, 120}
\definecolor{sgGreen} {RGB}{20, 180, 50}
\definecolor{revised}{rgb}{0,0,0.9}

\usepackage{enumitem}
\newcounter{descriptcount}
\renewcommand*\thedescriptcount{\arabic{descriptcount}}
\setlist[description]{leftmargin=0.25cm,labelindent=0.25cm}

\usepackage{mathptmx}

\usepackage{graphics}

\usepackage[authoryear,comma,longnamesfirst,sectionbib]{natbib}

\begin{document}

\title{Scalable  Collaborative Targeted  Learning  for High-Dimensional  Data}

\author{Cheng Ju, Susan Gruber, Samuel D. Lendle, Antoine Chambaz,\\ Jessica M. Franklin, Richard Wyss, Sebastian Schneeweiss, \\Mark J. van der Laan}

\date{}
\maketitle

\begin{abstract}
  
  Robust inference of  a low-dimensional parameter in  a large semi-parametric
  model relies on external estimators  of infinite-dimensional features of the
  distribution of  the data.  Typically, only  one of the latter  is optimized
  for the sake of constructing a well behaved estimator of the low-dimensional
  parameter of  interest.  Optimizing more  than one of  them for the  sake of
  achieving  a  better  bias-variance  trade-off  in  the  estimation  of  the
  parameter of interest  is the core idea driving the  general template of the
  collaborative targeted minimum loss-based estimation (C-TMLE) procedure.

  The  original implementation/instantiation  of  the C-TMLE  template can  be
  presented as a greedy forward stepwise  C-TMLE algorithm.  It does not scale
  well  when  the  number  $p$  of  covariates  increases  drastically.   This
  motivates the  introduction of  a novel implementation/instantiation  of the
  C-TMLE template where  the covariates are pre-ordered.   Its time complexity
  is  $\mathcal{O}(p)$  as  opposed  to  the  original  $\mathcal{O}(p^2)$,  a
  remarkable gain.  We propose two  pre-ordering strategies and suggest a rule
  of  thumb to  develop other  meaningful strategies.   Because it  is usually
  unclear a  priori which pre-ordering  strategy to choose, we  also introduce
  another implementation/instantiation called SL-C-TMLE algorithm that enables
  the data-driven choice of the better pre-ordering strategy given the problem
  at hand.  Its time complexity is $\mathcal{O}(p)$ as well.

  The computational  burden and relative performance of these algorithms were compared  in 
  simulation  studies involving
  fully synthetic  data or  partially synthetic  data based  on a  real world large
  electronic  health  database;  and  in 
  analyses of three real, large  electronic health databases.  In all analyses
  involving  electronic  health  databases,  the greedy  C-TMLE  algorithm  is
  unacceptably slow.  Simulation studies indicate our scalable C-TMLE and SL-C-TMLE
  algorithms work well. All C-TMLEs are publicly available in a Julia software package.
  
\end{abstract}

\section{Introduction}

The  general template  of collaborative  double robust  targeted minimum  loss-based
estimation  (C-\-TM\-LE;  ``C-TMLE  template''  for  short)  builds  upon  the
targeted      minimum      loss-based     estimation      (TMLE)      template
\citep{van2011targeted,van2010collaborative}.   Both   the  TMLE   and  C-TMLE
templates can  be viewed as meta-algorithms  which map a set  of user-supplied
choices/hyper-parameters ({  e.g.}, parameter  of interest,  loss function,
submodels) into a specific machine-learning  algorithm for estimation, that we
call an instantiation of the template. 

Constructing  a TMLE  or  a  C-TMLE   involves  the  estimation of  a
nuisance  parameter,   typically  an   infinite-dimensional  feature   of  the
distribution of the data. For a  vanilla TMLE estimator, the estimation of the
nuisance parameter  is addressed as  an independent statistical task.   In the
C-TMLE template, on the contrary, the  estimation of the nuisance parameter is
optimized to provide a better bias-variance  trade-off in the inference of the
targeted parameter.   The C-TMLE template  has been successfully applied  in a
variety of areas, from survival analysis \citep{stitelman2011targeted}, to the
study  of  gene  association  \citep{wang2011finding}  and  longitudinal  data
structures \citep{stitelman2010collaborative} to name just a few.

In    the    original    instantiation    of   the    C-TMLE    template    of
\citet{van2010collaborative},  that we  henceforth  call  ``the greedy  C-TMLE
algorithm'',  the estimation  of the  nuisance parameter  aiming for  a better
bias-variance trade-off  is conducted in  two steps.  First, a  greedy forward
stepwise selection procedure is implemented  to construct a nested sequence of
candidate estimators  of the nuisance parameter.   Second, cross-validation is
used to  select the candidate from  this sequence which minimizes  a criterion
that incorporates  a measure of  bias and variance  with respect to  (wrt) the
targeted    parameter   (the    algorithm   is    described      in
Section~\ref{sec:generalCTMLE}).  The authors show the greedy C-TMLE algorithm
exhibits superior relative performance in analyses of sparse data, at the cost
of  an increase  in time  complexity.   For instance,  in a  problem with  $p$
baseline  covariates,  one  would  construct and  select  from  $p$  candidate
estimators  of the  nuisance parameter,  yielding a  time complexity  of order
$\mathcal{O}(p^2)$.  Despite a criterion  for early termination, the algorithm
does  not scale  to large-scale  and high-dimensional  data. The  aim of  this
article  is to  develop novel  C-TMLE algorithms  that overcome  these serious
practical limitations without compromising  finite sample or asymptotic
performance. \\

We propose two  such ``scalable C-TMLE algorithms''.  They  replace the greedy
search at  each step by an  easily computed data adaptive  pre-ordering of the
candidate estimators of the nuisance parameter.  They include a data adaptive,
early  stopping   rule  that   further  reduces  computational   time  without
sacrificing statistical  performance.  In the aforementioned  problem with $p$
baseline covariates where  the time complexity of the  greedy C-TMLE algorithm
was  of order  $\mathcal{O}(p^2)$,  those  of the  two  novel scalable  C-TMLE
algorithms is of order $\mathcal{O}(p)$.

Because  one may  be  reluctant to  specify  a single  a priori    pre-ordering of  the
candidate estimators of the nuisance  parameter, we also introduce a SL-C-TMLE
algorithm. It selects the best pre-ordering  from a set of ordering strategies
by super  learning (SL)  \citep{van2007super}.  SL is  an example  of ensemble
learning  methodology which  builds a  meta-algorithm for  estimation out  of a
collection of individual, competing algorithms of estimation, relying on oracle properties of 
cross-validation.

We focus on the estimation of  the average (causal) treatment effect (ATE). It
is not hard  to generalize our scalable C-TMLE algorithms  to other estimation
problems.

The  performance of  the two  scalable  C-TMLE and  SL-C-TMLE algorithms  are
compared  with  those  of  competing,  well  established  estimation  methods:
G-computation  \citep{Robins86}, inverse  probability  of treatment  weighting
(IPTW) \citep{hernan_brumback_robins, Robins98}, augmented inverse probability
of   treatment  weighted   estimator  (A-IPTW)   \citep{Robins00a,  Robins00b,
  Robins00c}.   Results  from  unadjusted  regression estimation  of  a  point
treatment effect are
also provided to illustrate the level of bias due to confounding.\\

The article is organized  as follows.  Section~\ref{sec:background} introduces
the parameter  of interest and a  causal model for its  causal interpretation.
Section~\ref{sec:review}  describes an  instantiation  of  the TMLE  template.
Section~\ref{sec:generalCTMLE}  presents  the  C-TMLE  template  and  a  greedy
instantiation  of  it.   Section~\ref{sec:scalableCTMLE}  introduces  the  two
proposed  pre-ordered scalable  C-TMLE  algorithms,  and SL-C-TMLE  algorithm.
Sections~\ref{sec:sim}  and  \ref{subsec:sim:five}   present  the  results  of
simulation studies (based on fully  or partially synthetic data, respectively)
comparing the  C-TMLE and SL-C-TMLE  estimators with other  common estimators.
Section~\ref{sec:discussion} is  a closing  discussion. The  appendix presents
additional material: an  introduction to a Julia software  that implements all
the  proposed  C-TMLE algorithms;  a  brief  analysis of  their  computational
performance; the results of their application  to the analysis of three large
electronic health databases.

\section{The Average Treatment Effect Example}
\label{sec:background}

We consider the problem of estimating  the ATE in an observational study where
we  observe  on  each  experimental   unit:  a  collection  of  $p$  baseline
covariates, $W$; a binary treatment indicator, $A$; a binary or bounded continuous $(0,1)$-valued
outcome of  interest, $Y$.  We  use $O_i = (W_i,  A_i, Y_i)$ to  represent the
$i$-th  observation from  the unknown  observed data  distribution $P_0$,  and
assume that $O_{1}, \ldots, O_{n}$  are independent. The parameter of interest
is defined as
\begin{equation*}
  \Psi(P_0) = \E_0[\E_0(Y \mid A = 1, W) - \E_0(Y \mid A = 0, W)].  
\end{equation*}

The  ATE   enjoys  a  causal  interpretation under  the  non-parametric
structural equation model (NPSEM) given by:
\begin{equation*}
  \left\{%
    \begin{array}{l}
      W=f_W(U_W),\\
      A=f_A(W, U_A),\\
      Y=f_Y(A, W, U_Y),
    \end{array},\right.
\end{equation*}
where   $f_{W}$,  $f_{A}$   and  $f_{Y}$   are  deterministic   functions  and
$U_W, U_A,  U_Y$ are background  (exogenous) variables. The  potential outcome
under exposure level  $a \in \{0,1\}$ can be obtained  by substituting $a$ for
$A$ in  the third equality:  $Y_a =  f_Y(a, W, U_Y)$.   Note that $Y  = Y_{A}$
(this  is known  as the  ``consistency'' assumption).   If we  are willing  to
assume that {\em  (i)} $A$ is conditionally independent of  $(Y_1, Y_0)$ given
$W$ (this is  known as the ``no unmeasured confounders''  assumption) and {\em
  (ii)} $0 < P(A=1 \mid W) < 1$ almost everywhere (known as the ``positivity''
assumption), then parameter
$\Psi(P_{0})$ satisfies $\Psi(P_{0})=\E_{0}(Y_1-Y_0)$.\\

For  future use,  we  introduce  the propensity  score  (PS),  defined as  the
conditional probability of receiving treatment, and define $g_0(a,W) \equiv P_{0}(A=a \mid W)$
for  both  $a=0,1$.   We  also  introduce  the  conditional  mean of the outcome:
$\bar{Q}_0(A,W)= \E_{0}(Y \mid A,W)$. In  the remainder of this article, $g_n(a,W)$
and $\bar{Q}_n(A,W)$ denote estimators of $g_0(a,W)$ and $\bar{Q}_{0}(A,W)$.

\section{A TMLE Instantiation for ATE}
\label{sec:review}

We are  mainly interested in  double robust (DR) estimators  of $\Psi(P_{0})$.
An estimator of $\Psi(P_{0})$ is said to be  DR if it is consistent if either
$\bar{Q}_{0}$ or $g_{0}$ is consistently estimated.  In addition, an estimator
of $\Psi(P_{0})$  is said to be efficient  if it  satisfies a central  limit theorem
with a limit variance  which equals the second moment under  $P_{0}$ of the so
called efficient influence curve (EIC) at $P_{0}$.  The EIC for the ATE parameter is given by
\begin{equation*}
  D^*(\bar{Q}_0, g_{0})(O) = H_0(A,W) [Y - \bar{Q}_{0}(A,W)] + \bar{Q}_{0}(1, W) - 
  \bar{Q}_{0}(0,W) - \Psi(P_{0}),
\end{equation*}
where  $H_{0}(a,W) =  a /  g_{0}(1,W) -  (1-a) /  g_{0}(0,W)$ ($a=0,1)$.   The
notation is  slightly misleading because  there is more to  $\Psi(P_{0})$ than
$(\bar{Q}_{0},  g_{0})$  (namely,  the  marginal  distribution  of  $W$  under
$P_{0}$).  We nevertheless keep it that  way for brevity.  We refer the reader
to \citep{bickel1998efficient} for details about efficient influence curves.

More generally, for every valid distribution $P$ of $O=(W,A,Y)$ such that {\em
  (i)} the conditional expectation of  $Y$ given $(A,W)$ equals $\bar{Q}(A,W)$
and the conditional probability that $A=a$ given $W$ equals $g(a,W)$, and {\em
  (ii)} $0<g(1,W)<1$ almost surely, we denote
\begin{equation*}
  D^*(\bar{Q}, g)(O) = H_{g}(A,W) [Y - \bar{Q}(A,W)] + \bar{Q}(1, W) - 
  \bar{Q}(0,W) - \Psi(P),
\end{equation*}
where $H_{g}(a,W) = a / g(1,W) - (1-a) / g(0,W)$ ($a=0,1$).

The augmented  inverse probability of treatment weighted estimator    (A-IPTW,   or   so   called    ``double   robust   IPTW'';
\citet{robins1994estimation,robins2000marginal,van2003unified})    and    TMLE
\citep{van2006targeted, van2011targeted}  are two well studied  DR estimators.
Taking the  estimation of ATE as  an example, A-IPTW estimates  $\Psi(P_0)$ by
solving the  EIC equation directly.   For given estimators  $\bar{Q}_n$, $g_n$
and with 
\begin{equation}
  \label{eq:Hgn}
  H_{g_n}(a,W) = a/g_{n}(1,W) - (1-a) /g_{n}(0,W) \quad (a=0,1),
\end{equation}
solving (in $\psi$)
\begin{eqnarray*}
  0
  &=& \sum_{i=1}^{n}  H_{g_n}(A_i,W_i) [Y_i  - \bar{Q}_n(A_i,W_i)]  + \bar{Q}_n(1,
      W_i) - \bar{Q}_n(0,W_i) - \psi 
\end{eqnarray*}
yields the A-IPTW estimator
\begin{equation*}
  \label{eq:A-IPTW:one}
  \psi_n^{A-IPTW} = \sum_{i=1}^{n}H_{g_n}(A_i,W_i)  [Y_i - \bar{Q}_n(A_i,W_i)] +
  \bar{Q}_n(1, W_i) - \bar{Q}_n(0,W_i). 
\end{equation*}
  
A  substitution  (or   plug-in)  estimator  of  $\Psi(P_0)$   is  obtained  by
plugging-in  the  estimator   of  a  relevant  part   of  the  data-generating
distribution $P_{0}$ into the  mapping $\Psi$.  Substitution estimators belong
to the parameter space by definition, which is a desirable property.  The A-IPTW is
not a  substitution estimator  and can  suffer from it by sometimes producing estimates
outside of known bounds on the problem, such as probabilities or proportions greater than 1.   
On the  contrary, an
instantiation  of the  TMLE template  yields a  DR TMLE  estimator defined  by
substitution.  For  instance, a TMLE  estimator can  be can be  constructed by
applying  the  TMLE  algorithm  below   (which  corresponds  to  the  negative
$\log$-likelihood loss function and logistic fluctuation submodels).
\begin{enumerate}
\item {\bf Estimating $\bar{Q}_0$.}  Derive an initial estimator $\bar{Q}_n^0$
  of  $\bar{Q}_0$.   It  is  highly recommended  to  avoid  making  parametric
  assumptions,  as any  parametric model  is likely  mis-specified. Relying  on
  SL~\citep{van2007super} is a good option.
\item {\bf Estimating $g_0$.}  Derive an  estimator $g_n$ of $g_{0}$, The same
  recommendation as above  applies.
\item {\bf Building  the so called ``clever covariates''.}  For  $a=0,1$ and a
  generic $W$, define $H_n(a,W)$ as in \eqref{eq:Hgn}.
\item  {\bf   Targeting.}   Fit   the  logistic   regression  of   $Y_{i}$  on
  $H_n(A_i,W_i)$ with no  intercept, using $\logit(\bar{Q}_n^0(A_{i}, W_{i}))$
  as offset (an $i$-specific intercept).  This yields a minimum loss estimator
  $\epsilon_{n}$.    Update   the   initial   estimator   $\bar{Q}_n^0$   into
  $\bar{Q}_n^*$ given by
  \begin{equation*}
    \bar{Q}_n^*(A,W) = \expit\{\logit[\bar{Q}_n^0(A,W)] + \epsilon_n H_n(A,W)\}.
  \end{equation*}
\item {\bf Evaluating the parameter estimate}. Define
  \begin{equation}
    \label{eq:TMLE}
    \psi_n^{TMLE}   =    \frac{1}{n}   \sum_{i=1}^n    (\bar{Q}_n^*(1,W_i)   -
    \bar{Q}_n^*(0,W_i)). 
  \end{equation}
  As emphasized,  TMLE  is a \emph{substitution}
  estimator.
\end{enumerate}

The targeting step  aims to reduce bias in the  estimation of $\Psi(P_{0})$ by
enhancing the initial estimator derived  from $\bar{Q}_n^{0}$ and the marginal
empirical  distribution  of $W$  as  an  estimator  of its  counterpart  under
$P_{0}$.  The  fluctuation is  made in  such a  way that  the EIC  equation is
solved: $\sum D^*(\bar{Q}^*_n, g_n)(O_i)=0$.  Therefore, the TMLE estimator is
double   robust   and   (locally)  efficient   under   regularity   conditions
\citep{van2011targeted}.

Standard errors  and confidence intervals (CIs)  can be computed based  on the
variance of the  influence curve.  Proofs and technical  details are available
in the literature~\citep[][for instance]{van2006targeted,van2011targeted}.

In practice, bounded continuous outcomes and binary outcomes are fluctuated on
the  $\logit$   scale  to  ensure   that  bounds   on  the  model   space  are
respected~\citep{susan2010targeted}.

\section{The C-TMLE General Template and Its Greedy Instantiation for ATE}
\label{sec:generalCTMLE}


When  implementing an  instantiation of  the TMLE  template, one  relies on  a
single external estimate of the nuisance parameter, $g_{0}$ in the ATE example
(see Step~2  in Section~\ref{sec:review}).   In contrast, an  instantiation of
the  C-TMLE template  involves  constructing a  series  of nuisance  parameter
estimates  and corresponding  TMLE  estimators using  these  estimates in  the
targeting step.  

\subsection{The C-TMLE Template}
\label{subsec:C-TMLE}

When  the  ATE is  the  parameter  of interest,  the  C-TMLE  template can  be
summarized  recursively  like  this   (see  Algorithm~\ref{algo:ctmle}  for  a
high-level      algorithmic     presentation).       One     first      builds
$(g_{n,0}, \bar{Q}_{n}^{0}=\bar{Q}_{n,0},  \bar{Q}_{n,0}^{*})$ where $g_{n,0}$
is           an          estimator           of          $g_{0}$           and
$\bar{Q}_{n}^{0}=\bar{Q}_{n,0},    \bar{Q}_{n,0}^{*}$   are    estimators   of
$\bar{Q}_{0}$, the latter being targeted  toward the parameter of interest for
instance  as   in  Section~\ref{sec:review}.   Given  the   previous  triplets
$(g_{n,0},    \bar{Q}_{n}^{0}=\bar{Q}_{n,0},    \bar{Q}_{n,0}^{*}),    \ldots,
(g_{n,k-1}, \bar{Q}_{n,k-1}, \bar{Q}_{n,k-1}^{*})$ where, by construction, the
empirical  loss  of  each  $\bar{Q}_{n,\ell}^{*}$  is  smaller  than  that  of
$\bar{Q}_{n,\ell-1}^{*}$,  one  needs to  generate  the  next triplet  in  the
sequence.  The  current initial estimator  of $\bar{Q}_{0}$ at  the $(k+1)$-th
step is set  at $\bar{Q}_{n,k}=\bar{Q}_{n,k-1}$ ({\em i.e.}, the  same as that
from triplet $(g_{n,k-1}, \bar{Q}_{n,k-1}, \bar{Q}_{n,k-1}^*)$).  One then has
a set of moves to create candidates $g_{n,k}^j$ updating $g_{n,k-1}$ with move
$j$ ({\em e.g.}, adding $j$-th covariate), providing better empirical fit than
$g_{n,k-1}$  and   yielding  the  corresponding   $\bar{Q}_{n,k}^{j,*}$  using
$\bar{Q}_{n,k}=\bar{Q}_{n,k-1}$ as  initial.  The candidate with  the smallest
empirical loss  is $(g_{n,k},  \bar{Q}_{n,k}, \bar{Q}_{n,k}^{*})$.   Two cases
arise: if the  empirical loss of the candidate  $\bar{Q}_{n,k}^{*}$ is smaller
than  that of  $\bar{Q}_{n,k-1}^{*}$, then  one has  derived the  next triplet
$(g_{n,k},  \bar{Q}_{n,k}=\bar{Q}_{n,k-1}, \bar{Q}_{n,k}^{*})$;  otherwise, in
our sequence,  one updates the initial  $\bar{Q}_{n,k}=\bar{Q}_{n,k-1}^{*}$ to
the $\bar{Q}_{n,k-1}^{*}$  in the last triplet,  and one repeats the  above to
generate  $(g_{n,k}, \bar{Q}_{n,k},  \bar{Q}_{n,k}^{*})$  -- since  it is  now
guaranteed that the empirical loss of $\bar{Q}_{n,k}^{*}$ is smaller that that
of   $\bar{Q}_{n,k-1}^{*}$,  one   always  gets   the  desired   next  element
$(g_{n,k}, \bar{Q}_{n,k}, \bar{Q}_{n,k}^{*})$.

In  the  original  greedy C-TMLE  algorithm~\citep{van2010collaborative},  the
successive nuisance parameter  estimates are based on  a data-adaptive forward
stepwise search that optimizes a goodness-of-fit criterion at each step.  Each
of them then yields a specific, candidate TMLE.  Finally, the C-TMLE
is  defined as  that  candidate  that optimizes  a  cross-validated
version of the criterion.  The C-TMLE  inherits all the properties of
a vanilla  TMLE estimator~\citep{van2010collaborative}.   It is  double robust
and asymptotically efficient under appropriate regularity conditions.

%
%
%

\begin{algorithm}[H]
  \caption{General Template of C-TMLE}
  \label{algo:ctmle}
  \begin{algorithmic}[1]
    \State{Construct an initial estimate $\bar{Q}_n^0$ for $\bar{Q}_0.$ }

    \State{Create candidate $\bar{Q}_{n,k}^{*}$,  using different estimates of
      treatment   mechanism  $g_0$,   such  that   the  empirical   losses  of
      $\bar{Q}_{n,k}^{*}$  and $g_{n,k}$  are decreasing  in $k$.   The greedy
      C-TMLE algorithm uses a forward greedy selection algorithm.}

    \State{Select  the best  candidate  $\bar{Q}_n^{*} =  \bar{Q}_{n,k_n}^{*}$
      using loss-based cross-validation, with the same loss function as in the
      TMLE targeting step.}
  \end{algorithmic}
\end{algorithm}

In  Step~1  of  Algorithm~\ref{algo:ctmle},  we   recommend  using  SL  as described further in
Section~\ref{sec:review}.  Step~2  will be commented  on in the  next section.
In  Step~3,  the best  candidate  is  selected  based on  the  cross-validated
penalized log-likelihood and indexed by
\begin{equation*}
  k_n = \argmin_{k} \left\{\cvRSS + \cvVar_k +  n \times \cvBias_k^2\right\}
\end{equation*}
where
\begin{eqnarray*}
  \cvRSS_k 
  &=& \sum_{v=1}^{V} \sum_{i \in \Val(v)} (Y_i - \bar{Q}_{n,k}^*(P^0_{nv})(W_i, A_i))^2,\\
  \cvVar_k 
  &=&  \sum_{v=1}^{V}\sum_{i  \in \Val(v)}D^{*2}  (\bar{Q}_{n,k}^*(P^0_{nv}),
      g_{n,k}(P_n))(O_i),\\ 
  \cvBias_k 
  &=&     \frac{1}{V}      \sum_{v=1}^{V}\Psi(\bar{Q}_{n,k}^*(P_{nv}^0))     -
      \Psi(\bar{Q}_{n,k}^*(P_n)).
\end{eqnarray*}
In the above display, $\Val(v)$ is the set of indices of observations used for
validation in the  $v$-th fold, $P_{nv}^{0}$ is the  empirical distribution of
the  observations indexed  by $i  \not\in \Val(v)$,  $P_{n}$ is  the empirical
distribution  of  the  whole  data  set,  and  $Z(P_{nv}^{0})$  (respectively,
$Z(P_{n})$)  means  that  $Z$  is  fitted  using  $P_{nv}^{0}$  (respectively,
$P_{n}$).  The penalization terms $\cvVar_{k}$ and $\cvBias_{k}$ robustify the
finite    sample   performance    when    the    positivity   assumption    is
violated~\citep{van2010collaborative}.

To  achieve  collaborative  double  robustness,  the  sequence  of  estimators
$(g_{n,k} : k)$ should be arranged in  such a way that the bias is monotonically decreasing while the
variance is monotonically increasing such that $g_{n,k}$  converges (in $k$)  to a
consistent estimator of $g_0$~\citep{van2011targeted}.  One could for instance
rely  on a  nested sequence  of models,  see Section~\ref{subsec:origCTMLE}.   By
doing    so,   the    empirical   fit    for   $g_{0}$    improves   as    $k$
increases~\citep{van2011targeted,gruber2010application}.

\citet{porter2011relative} discuss and  compare TMLE and C-TMLE  with other DR
estimators, including A-IPTW.

\subsection{The Greedy C-TMLE Algorithm}
\label{subsec:origCTMLE}

We refer to the original instantiation of the C-TMLE
template as the greedy C-TMLE algorithm. It uses  a  forward  selection  algorithm  to  build  the  sequence  of
estimators of $g_{0}$ as a  nested sequence of treatment models.  Let us
describe it  in the  case that $W$  consists of $p$  covariates. For  $k=0$, a
one-dimensional  logistic model  with only  an intercept  is used  to estimate
$g_{0}$.   Recursively, the  $(k+1)$th model  is  derived by  adding one  more
covariate  from $W$  to the  $k$th logistic  model.  The  chosen covariate  is
selected from the set of covariates in $W$ that have not been selected so far.

More  specifically,  one  begins  with  the intercept  model  for  $g_{0}$  to
construct $g_{n,0}$  then a  first fluctuation  covariate $H_{g_{n,0}}$  as in
\eqref{eq:Hgn}, which is used in turn  to create the first candidate estimator
$\bar{Q}_{n,0}^{*}$    based    on    $\bar{Q}_{n,0}$.     Namely,    denoting
$g_{n,0}(1 \mid W) = P_{n}(A=1)$ and $g_{n,0}(0 \mid W) = P_{n}(A=0)$, we set
\begin{eqnarray}
  \label{eq:clever}
  H_{g_{n,k}}(a,W) 
  &=& a/g_{n,k}(1 \mid W) - (1-a)/g_{n,k} (0 \mid W),\\ 
  \label{eq:fluct}
  \logit(\bar{Q}_{n,k}^{*}(a,W))
  &=& \logit(\bar{Q}_{n,k}(a,W)) + \epsilon_{k}H_{g_{n,k}}(a,W) \quad (a=0,1) 
\end{eqnarray}
where $k=0$.  Here $\epsilon_{k}$ is fitted by a logistic regression of $Y$ on
$H_{g_{n,k}}(A,W)$ with  offset $\bar{Q}_{n,k}(A,W)$,  and $\bar{Q}_{n,1}^{*}$
is the first  candidate TMLE.  We denote $\xL_{0}$ its  empirical loss wrt the
negative $\log$-likelihood function $\xL$.

We   proceed    recursively.    Assume   that   we    have   already   derived
$\bar{Q}_{n,0}^{*},  \ldots$, $\bar{Q}_{n,k-1}^{*}$,  and  denote the  initial
estimator used  in the  last TMLE $\bar{Q}_{n,k-1}^*$  with $\bar{Q}_{n,k-1}$.
The $(k+1)$-th estimator $g_{n,k}$ of $g_{0}$  is based on a larger model than
that we  yielded $g_{n,k-1}$. It contains  the intercept and the  same $(k-1)$
covariates  as  the  previous  model  fit  $g_{n,k-1}$,  with  one  additional
covariate.  Each covariate $W_j$ ($1 \leq j  \leq p$ such that $W_{j}$ has not
been selected yet) is considered in  turn for inclusion in the model, yielding
a  update $g_{n,k}^{j}$  of $g_{n,k-1}$,  which implies  corresponding updates
$H_{g_{n,k}}^{j}$ and $\bar{Q}_{n,k}^{j,*}$  as in the above  display.  A best
update   $\bar{Q}_{n,k}^{*}$  is   selected   among   the  candidate   updates
$\bar{Q}_{n,k}^{1,*}, \ldots, \bar{Q}_{n,k}^{p,*}$ by minimizing the empirical
loss   wrt   $\xL$.    Its   empirical  loss   is   denoted   $\xL_{k}$.    If
$\xL_{k}  \leq  \xL_{k-1}$,  then  this  $\bar{Q}_{n,k}^*$  defines  the  next
fluctuation  in  our  sequence,  with corresponding  initial  estimator  still
$\bar{Q}_{n,k}=\bar{Q}_{n,k-1}$,   the   same   as    that   used   to   build
$\bar{Q}_{n,k-1}^*$. We can now move on to the next step.  Otherwise, we reset
the initial estimator $\bar{Q}_{n,k-1}$  to $\bar{Q}_{n,k-1}^*$ and repeat the
above   procedure:   {\em   i.e.},    we   compute   the   candidate   updates
$\bar{Q}_{n,k}^{j,*}$ again  for this  new initial  estimator, and  select the
best   choice   $\bar{Q}_{n,k}^*$.    Due   to  the   initial   estimator   in
$\bar{Q}_{n,k}^*$ being $\bar{Q}_{n,k-1}^*$, it is now guaranteed that the new
$\xL_{k}$ is smaller than $\xL_{k-1}$, thereby providing us with our next TMLE
$\bar{Q}_{n,k}^*$ in our sequence.

This  forward stepwise  procedure is  carried  out recursively  until all  $p$
covariates  have  been  incorporated  into  the model  for  $g_{0}$.   In  the
discussed  setting, choosing  the  first covariate  requires $p$  comparisons,
choosing the second  covariate requires $(p-1)$ comparisons and  so on, making
the time complexity of this algorithm $\mathcal{O}(p^2)$.

Once all  candidates $\bar{Q}_{n,0}^{*}, \ldots, \bar{Q}_{n,k}^{*}$  have been
constructed,  cross-validation  is  used  to  select  the  optimal  number  of
covariates to include  in the model for $g_{0}$.  For  more concrete examples,
we                                                                       refer
to~\citep{van2010collaborative,van2011targeted}. \citet{gruber2010application}
proposes   several  variations   on   the  forward   greedy  stepwise   C-TMLE
algorithm. The variations  did not improve performance  in simulation studies.
In  this article,  the greedy  C-TMLE algorithm  is defined  by the  procedure
described above.

\section{Scalable C-TMLE Algorithms}
\label{sec:scalableCTMLE}

Now that  we have introduced the  background on C-TMLE, we  will now introduce
our  scalable C-TMLE  algorithm.  Section~\ref{subsec:outline}  summarizes the
philosophy of the scalable C-TMLE algorithm, which hinges on a data adaptively
determined      pre-ordering       of      the       baseline      covariates.
Sections~\ref{subsec:logistic}   and   \ref{subsec:corr}  present   two   such
pre-ordering  strategies.    Section~\ref{subsec:pre:order:discuss}  discusses
what   properties   a   pre-ordering  strategy   should   satisfy.    Finally,
Section~\ref{subsec:SL-CTMLE}  proposes a  discrete Super  Learner-based model
selection procedure to select among a set of scalable C-TMLE estimators, which
is itself a scalable C-TMLE algorithm.

\subsection{Outline}
\label{subsec:outline}

As we  have seen in  the previous section, the  time complexity of  the greedy
C-TMLE algorithm  is $\mathcal{O}(p^2)$ when  the number of  covariates equals
$p$. This is  unsatisfactory for large scale and  high-dimensional data, which
is an increasingly common situation in health care research.  For example, the
high-dimensional  propensity score  (hdPS) algorithm  is a  method to  extract
information from electronic medical claims data that produces hundreds or even
thousands  of  candidate covariates,  increasing  the  dimension of  the  data
dramatically~\citep{schneeweiss2009high}.

In order to make it possible to  apply C-TMLE algorithms to such data sets, we
propose to  add a new pre-ordering  procedure after the initial  estimation of
$\bar{Q}_{0}$  and   before  the   stepwise  construction  of   the  candidate
$\bar{Q}_{n,k}^{*}$, $k=0,\ldots$.  We present  two pre-ordering procedures in
Sections~\ref{subsec:logistic} and \ref{subsec:corr}.  By imposing an ordering
over  the covariates only
one covariate is eligible for inclusion in the PS model at each step when constructing
the next candidate TMLE in the sequence, $\bar{Q}_{n,k}^{*}$.  Thus,
 the new  C-TMLE algorithm  overcomes the  computational
issue.

Once an ordering over the covariates has  been established, we add them one by
one to the model used to  estimate $g_{0}$, starting from the intercept model.
Suppose  that we  are adding  the $k$th  covariate; we  obtain a  new estimate
$g_{n,k}$   of    $g_{0}$;   we   define    a   new   clever    covariate   as
in~\eqref{eq:clever};   we    fluctuate   the   current    initial   estimator
$\bar{Q}_{n}^{k}$  as  in~\eqref{eq:fluct};  we evaluate  the  empirical  loss
$\xL_{k}$  wrt  $\xL$  of  the  resulting  candidate  $\bar{Q}_{n,k}^{*}$.  If
$\xL_{k}  \leq \xL_{k-1}$,  then  we move  on to  adding  the next  covariate;
otherwise,  the  current  initial  estimate  $\bar{Q}_{n,k}$  is  replaced  by
$\bar{Q}_{n,k-1}^{*}$ and we restart over  adding the $k$th covariate. This approach
 guarantees that $\xL_{k} \leq \xL_{k-1}$.

Finally,  we  use   cross-validation  to  select  the   best  candidate  among
$\bar{Q}_{n,0}^{*}, \ldots$,  $\bar{Q}_{n,p}^{*}$ in terms  of cross-validated
loss wrt $\xL$.



\subsection{Logistic Pre-Ordering Strategy}
\label{subsec:logistic}

The logistic  pre-ordering procedure  is similar  to the  second round  of the
greedy C-TMLE algorithm.   However, instead of selecting  one single covariate
before going on, we use the empirical losses wrt $\xL$ to order the covariates
by  their ability  to  reduce  bias.  More  specifically,  for each  covariate
$W_{k}$  ($1 \leq  k \leq  p$),  we construct  an estimator  $g_{n,k}$ of  the
conditional distribution of  $A$ given $W_{k}$ only (one might  also add $W_k$
to  a   fixed  baseline   model);  we   define  a   clever  covariate   as  in
\eqref{eq:clever}  using  $g_{n,k}$  and  fluctuate  $\bar{Q}_{n}^{0}$  as  in
\eqref{eq:fluct};   we   compute  the   empirical   loss   of  the   resulting
$\bar{Q}_{n,k}^{*}$ wrt  $\xL$, yielding  $\xL_{k}$.  Finally,  the covariates
are ranked by increasing values of  the empirical loss.  This is summarized in
Algorithm~\ref{algo:logist}.

\begin{algorithm}[H]
  \caption{Logistic Pre-Ordering Algorithm}
  \label{algo:logist}
  \begin{algorithmic}[1]
    \For{each covariate $W_k$ in $W$}

    \State{Construct an estimator $g_{n,k}$ of  $g_{0}$ using a logistic model
      with $W_{k}$ as predictor.}

    \State{Define    a   clever    covariate   $H_{g_{n,k}}(A,W_k)$    as   in
      \eqref{eq:clever}.}

    \State {Fit  $\epsilon_k$ by  regressing $Y$ on  $H_{g_{n,k}}(A,W_k)$ with
      offset $\bar{Q}_{n}^{0} (A,W)$.}

    \State{Define $\bar{Q}_{n,k}^{*}$ as in \eqref{eq:fluct}.}

    \State{Compute the empirical loss $\xL_{k}$ wrt $\xL$.}

    \EndFor 

    \State{Rank the covariates by increasing $\xL_k$.}
  \end{algorithmic}
\end{algorithm}

\subsection{Partial Correlation Pre-Ordering Strategy}
\label{subsec:corr}

In the  greedy C-TMLE  algorithm described  in Section~\ref{subsec:origCTMLE},
once  $k$  covariates  have  already  been selected,  the  $(k+1)$th  is  that
remaining covariate which provides the largest reduction in the empirical loss
wrt $\xL$. Intuitively, the $(k+1)$th covariate  is the one that best explains
the residual between $Y$ and  {\em the current} $\bar{Q}_{n}^{0}$.  Drawing on
this  idea,  the partial  correlation  pre-ordering  procedure ranks  the  $p$
covariates based on  how each of them is correlated  with the residual between
$Y$ and  {\em the initial}  $\bar{Q}_{n}^{0}$ within  strata of $A$.   This second
strategy is less computationally demanding than the previous one because there
is  no need  to fit  any  regression models,  merely to  estimate $p$  partial
correlation coefficients.

Let $\rho(X_{1},  X_{2})$ denote  the Pearson correlation  coefficient between
$X_{1}$    and    $X_{2}$.     Recall    that    the    partial    correlation
$\rho(X_{1},X_{2}|X_{3})$ between $X_{1}$ and $X_{2}$ given $X_{3}$ is defined
as  the   correlation  coefficient  between  the   residuals  $R_{X_{1}}$  and
$R_{X_{2}}$ resulting from the linear regression  of $X_{1}$ on $X_{3}$ and of
$X_{2}$  on  $X_{3}$,   respectively~\citep{hair2006multivariate}.   For  each
$1 \leq k \leq p$, we introduce $R=Y-\bar{Q}_{n}^{0} (A,W)$,
\begin{equation*}
  \rho(R,W_k | A) = \frac{\rho(R, W_k) -\rho(R, A) \times \rho(W_k,
    A)}{\sqrt{(1 - \rho(R, A)^2)(1 - \rho(W_k, A)^2)}}.
\end{equation*}
The   partial    correlation   pre-ordering   strategy   is    summarized   in
Algorithm~\ref{algo:corr}.

\begin{algorithm}[H]
  \caption{Partial Correlation Pre-Ordering Algorithm}
  \label{algo:corr}
  \begin{algorithmic}[1]

    \For{each covariate $W_k$ in $W$}

    \State{Estimate  the partial  correlation coefficient  $\rho(R, W_k  | A)$
      between $R=(Y - \bar{Q}_{n}^{0} (A,W))$ and $W_k$ given $A$.}
    \EndFor

    \State{Rank the covariates based on the absolute value of the estimates of
      the partial correlation coefficients.}
  \end{algorithmic}
\end{algorithm}

\subsection{Discussion of the Design of Pre-ordering}
\label{subsec:pre:order:discuss}

Sections~\ref{subsec:logistic} and \ref{subsec:corr}  proposed two pre-ordering
strategies. In general, a rule of  thumb for designing a pre-ordering strategy
is to rank the covariates based on the impact of each in reducing the residual
bias in the target parameter which results from
the initial estimator $\bar{Q}_{n}^{0}$ of  $\bar{Q}_{0}$.  In this light, the
logistic ordering  of Section~\ref{subsec:logistic}  uses TMLE to  reflect the
importance of  each variable wrt  its potential  to reduce residual  bias. The
partial correlation ordering of Section~\ref{subsec:corr} ranks the covariates
according to  the partial correlation of  residual of the initial  fit and the
covariates, conditional on treatment.

Because the rule  of thumb considers each covariate in  turn separately, it is
particularly relevant when the covariates are not too dependent.  For example,
consider  the extreme  case where  two or  more of  the covariates  are highly
correlated and can greatly explain the  residual bias in the target parameter.
In this scenario, these dependent covariates would {\em all} be ranked towards
the front  of the ordering.  However,  after adjusting for {\em  one} of them,
the others  would typically be  much less  helpful for reducing  the remaining
bias.   This  redundancy may  harm  the  estimation.   In  cases where  it  is
computationally  feasible, this  problem can  be avoided  by using  the greedy
search strategy, but many other intermediate strategies can be pursued as well.  

\subsection{Super Learner-Based C-TMLE Algorithm}
\label{subsec:SL-CTMLE}

Here,  we explain  how to  combine several  C-TMLE algorithms  into one.   The
combination is based  on a Super Learner (SL).  Super  learning is an ensemble
machine learning approach that relies on cross-validation.  It has been proven
that a  SL selector can perform  asymptotically as well as  an oracle selector
under mild assumptions \citep{van2007super,van2003unified,vaart2006oracle}.

As hinted at above, a SL-C-TMLE algorithm  is an instantiation of an extension of
the C-TMLE template.  It builds upon several competing C-TMLE algorithms, each
relying on different  strategies to construct a sequence of  estimators of the
nuisance  parameter.  A  SL-C-TMLE algorithm  can  be designed  to select  the
single best strategy (discrete SL-C-TMLE algorithm), or an optimal combination
thereof  (ensemble SL-C-TMLE  algorithm).  A  SL-C-TMLE algorithm  can include
both  greedy  search  and  pre-ordering methods.   A  SL-C-TMLE  algorithm  is
scalable if all of the candidate C-TMLE algorithms in the library are scalable
themselves.

We focus on a scalable discrete SL-C-TMLE algorithm that uses cross-validation
to   choose  among   candidate  scalable   (pre-ordered)  C-TMLE   algorithms.
Algorithm~\ref{algo:SL:ctmle}  describes  its  steps.    Note  that  a  single
cross-validation procedure is  used to select both the  ordering procedure $m$
and the  number of  covariates $k$ included  in the PS  model.  It  is because
computational  time  {\em is}  an  issue  that we  do  not  rely on  a  nested
cross-validation procedure to select $k$ for each pre-ordering strategy $m$.

\begin{algorithm}[H]
  \caption{Super Learner C-TMLE Algorithm}
  \label{algo:SL:ctmle}
  \begin{algorithmic}[1]
    \State{Define $M$ covariates pre-ordering strategies yielding $M$ C-TMLE algorithms} 

    \For{each pre-ordering strategy $m$} 

    \State{Follow  step~2 of  Algorithm~\ref{algo:ctmle}  to create  candidate
      $\bar{Q}_{n, m, k}^{*}$ for the $m$-th strategy.}

    \EndFor

    \State{The  best  candidate  $\bar{Q}_n^{*}$   is  the  minimizer  of  the
      cross-validated losses of $\bar{Q}_{n, m, k}^{*}$ across all the $(m,k)$
      combinations.}
  \end{algorithmic}
\end{algorithm}

The time complexity of the SL-C-TMLE algorithm is of the same order as that of
the  most  complex C-TMLE  algorithm  considered.   So, if  only  pre-ordering
strategies of order $\mathcal{O}(p)$ are  considered, then the time complexity
of the SL-C-TMLE  algorithm is $\mathcal{O}(p)$ as well.  Given  a constant number
of user-supplied strategies, the SL-C-TMLE  algorithm remains scalable, with a
processing time that is approximately equal
to the sum of the times for each strategy.\\
 

We  compare the  pre-ordered C-TMLE  algorithms and  SL-C-TMLE algorithm  with
greedy C-TMLE algorithm and other common methods in Sections~\ref{sec:sim} and
Appendix \ref{sec:analyses}.





\section{Simulation Studies on Fully Synthetic Data}
\label{sec:sim}

%
%

We carried out four Monte-Carlo  simulation studies to investigate and compare
the performance  of G-computation  (that we call  MLE), IPTW,  A-IPTW, greedy
C-TMLE algorithm and scalable C-TMLE algorithms to estimate the ATE parameter.
For  each study,  we  generated $N  =  1,000$ Monte-Carlo  data  sets of  size
$n =  1,000$.  Propensity score  estimates were  truncated to fall  within the
range $[0.025, 0.975]$ for all estimators.

Denoting $\bar{Q}_{n}^{0}$ and $g_{n}$ two initial estimators of $\bar{Q}_{0}$
and $g_{0}$, the unadjusted, G-computation/MLE, and IPTW estimators of the ATE
parameter are given by \eqref{eq:unadj}, \eqref{eq:Gcomp} and \eqref{eq:IPTW}:
\begin{eqnarray}
  \label{eq:unadj}
  \psi_n^{unadj} 
  &=&  \displaystyle  \frac{\sum_{i=1}^n I(A_i = 1)Y_i} {\sum_{i=1}^nI(A_i = 1)}    
      - \frac{\sum_{i=1}^n I(A_i = 0)Y_i} {\sum_{i=1}^nI(A_i = 0)},\\
  \label{eq:Gcomp}
  \psi_n^{MLE} 
  &=& \frac{1}{n} \displaystyle \sum_{i=1}^n [Q_n^0(1,W_i) - Q_n^0(0,W_i)],\\
  \label{eq:IPTW}
  \psi_n^{IPTW} 
  &=& \frac{1}{n} \displaystyle  \sum_{i=1}^n \left[I(A_i=1) - I(A_i=0)\right]
      \frac{Y_i}{g_n(A_i,W_i)}, \\
  \notag
  \psi_n^{A-IPTW} 
  &=&
      \frac{1}{n}   \displaystyle   \sum_{i=1}^n   \frac   {\left[I(A_i=1)   -
      I(A_i=0)\right]}{g_n(A_i \mid W_i)} (Y_i-Q^0_n(W_i,A_i)) \\
  \label{eq:A-IPTW}
  && +\frac{1}{n}\sum_{i=1}^n (Q^0_n(1,W_i)-Q^0_n(0,W_i)).
\end{eqnarray}
The A-IPTW and TMLE estimators  were presented in Section~\ref{sec:review}. The
estimators yielded by the C-TMLE  and scalable C-TMLE algorithms were presented
in Section~\ref{sec:generalCTMLE},
\ref{subsec:origCTMLE} and \ref{sec:scalableCTMLE}.\\

For all  simulation studies, $g_0$  was estimated using a  correctly specified
main  terms logistic  regression  model. Propensity  scores incorporated  into
IPTW, A-IPTW, and TMLE were based on  the full treatment model for $g_0$.  The
simulation studies  of Sections~\ref{subsec:sim:one}  and \ref{subsec:sim:two}
illustrate  the relative  performance  of the  estimators  in scenarios  with
highly correlated covariates.   These two scenarios are by far the
most challenging settings for the  greedy C-TMLE and
scalable    C-TMLE    algorithms.    The    simulation    studies    of
Section~\ref{subsec:sim:three}   and   \ref{subsec:sim:four}  illustrate 
performance in situations where instrumental variables (covariates predictive
of the treatment  but not of the  outcome) are included in the  true PS model.
In these two  scenarios, greedy C-TMLE and our scalable  C-TMLEs are
expected  to perform  better,  if  not much  better,  than  other widely  used
doubly-robust methods.


\subsection{Simulation Study 1: Low-dimensional, highly correlated covariates}
\label{subsec:sim:one}

In the first simulation study, data  were simulated based on a data generating
distribution published  by \citet{freedman2008weighting} and  further analyzed
by \citet{petersen2012diagnosing}.  A pair  of correlated, multivariate normal
baseline       covariates       $(W_1,W_2)$        is       generated       as
$(W_1,  W_2)  \sim  N(\mu,  \Sigma)$  where  $\mu_1 =  0.5,  \mu_2  =  1$  and
$\Sigma = \begin{bmatrix} 2 & 1 \\ 1 & 1
\end{bmatrix}$.  The PS is given by
$$P_{0}(A = 1 \mid W) = g_0(1 \mid W) = \expit(0.5 + 0.25 W_{1} + 0.75 W_{2})$$
(this is a  slight modification of the mechanism in  the original paper, which
used  a probit  model  to  generate treatment).   The  outcome is  continuous,
$Y = \bar{Q}_0(A,W)  + \epsilon$, with $\epsilon \sim  N(0,1)$ (independent of
$A,W$) and $\bar{Q}_0(A,W) = 1 + A + W_1 + 2W_2.$
The true value of the target parameter is $\psi_0 = 1$.  

Note that {\em (i)} the two baseline covariates are highly correlated and {\em
  (ii)}  the  choice of  $g_{0}$  yields  practical  (near) violation  of  the
positivity assumption. 

Each  of  the  estimators  involving   the  estimation  of  $\bar{Q}_{0}$  was
implemented twice, using or not a  correctly specified model to estimate $Q_0$
(the mis-specified model is  a linear regression model of $Y$  on $A$ and $W_1$
only).


\begin{table}[H]
  \centering
  \caption{Simulation study 1.  Performance  of the various estimators across
    1000 simulated data sets of sample size 1000.}
    \label{table:Freedman2008}
    \scalebox{0.8}{
  \begin{tabular}{l|rrr|rrr}
    \hline & \multicolumn{3}{c}{\textbf{correct $\bar{Q}$}} &
    \multicolumn{3}{c}{\textbf{mis-specified $\bar{Q}$}} \\ & bias ($10^{-3}$) & se ($10^{-2}$) & MSE ($10^{-3}$) &
    bias ($10^{-3}$) & se ($10^{-2}$) & MSE ($10^{-3}$) \\
    \hline
    unadj           & 2766.8 & 22.6 & 7706.3 & 2766.8 & 22.61 & 7706.3 \\ 
    A\_IPTW         & 0.7 &  9.54 & 9.1 &   10.8 & 13.52 & 18.4 \\ 
    IPTW            & 75.9 & 34.91 & 127.5 &75.9 & 34.91 & 127.5 \\ 
    MLE             &  1.0 & 8.20 & 6.7 & 699.4 & 13.96 & 508.6 \\ 
    TMLE            & 0.6 & 9.55 & 9.1 &    1.3 & 11.05 & 12.2 \\ 
    greedy C-TMLE   & 0.8 & 8.91 & 7.9 &    0.4 & 10.41 & 10.8 \\ 
    logRank C-TMLE  & 0.1 & 8.94 & 8.0 &    0.4 & 10.41 & 10.8 \\ 
    partRank C-TMLE & 0.3 & 8.94 & 8.0 &    0.4 & 10.41 & 10.8 \\ 
    SL-C-TMLE       & 0.1 & 9.07 & 8.2 &    0.4 & 10.41 & 10.8 \\ 
    \hline
  \end{tabular}
  }
\end{table}

\begin{figure}[H]
  \centering
  \begin{subfigure}[t]{0.4\textwidth}
    \includegraphics[width=\textwidth]{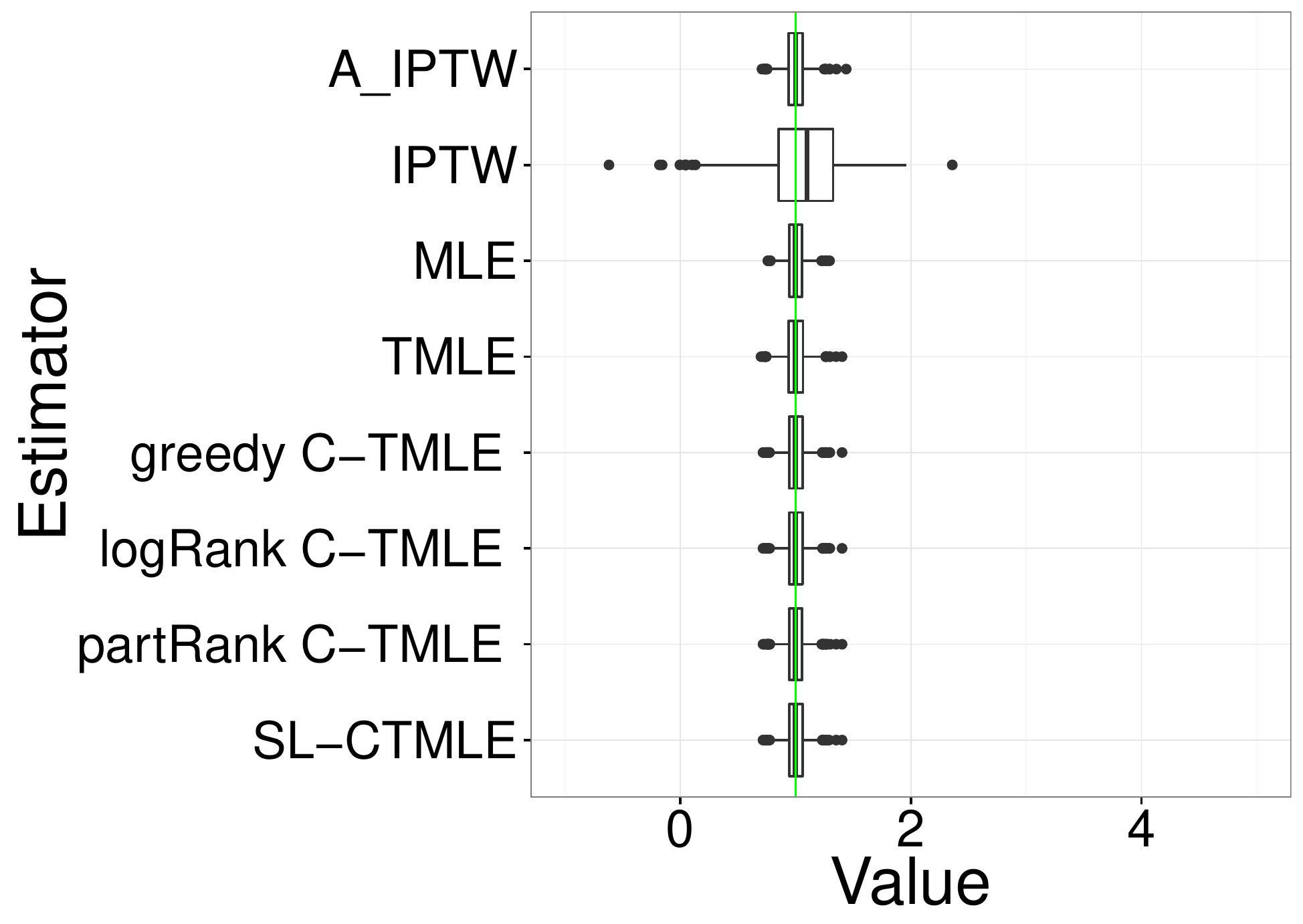}
    \caption{Well specified model for $\bar{Q_{0}}$.}
    \label{fig:Freedman2008-correct}
  \end{subfigure}
  \hspace{1cm}
  \begin{subfigure}[t]{0.4\textwidth}
    \includegraphics[width=\textwidth]{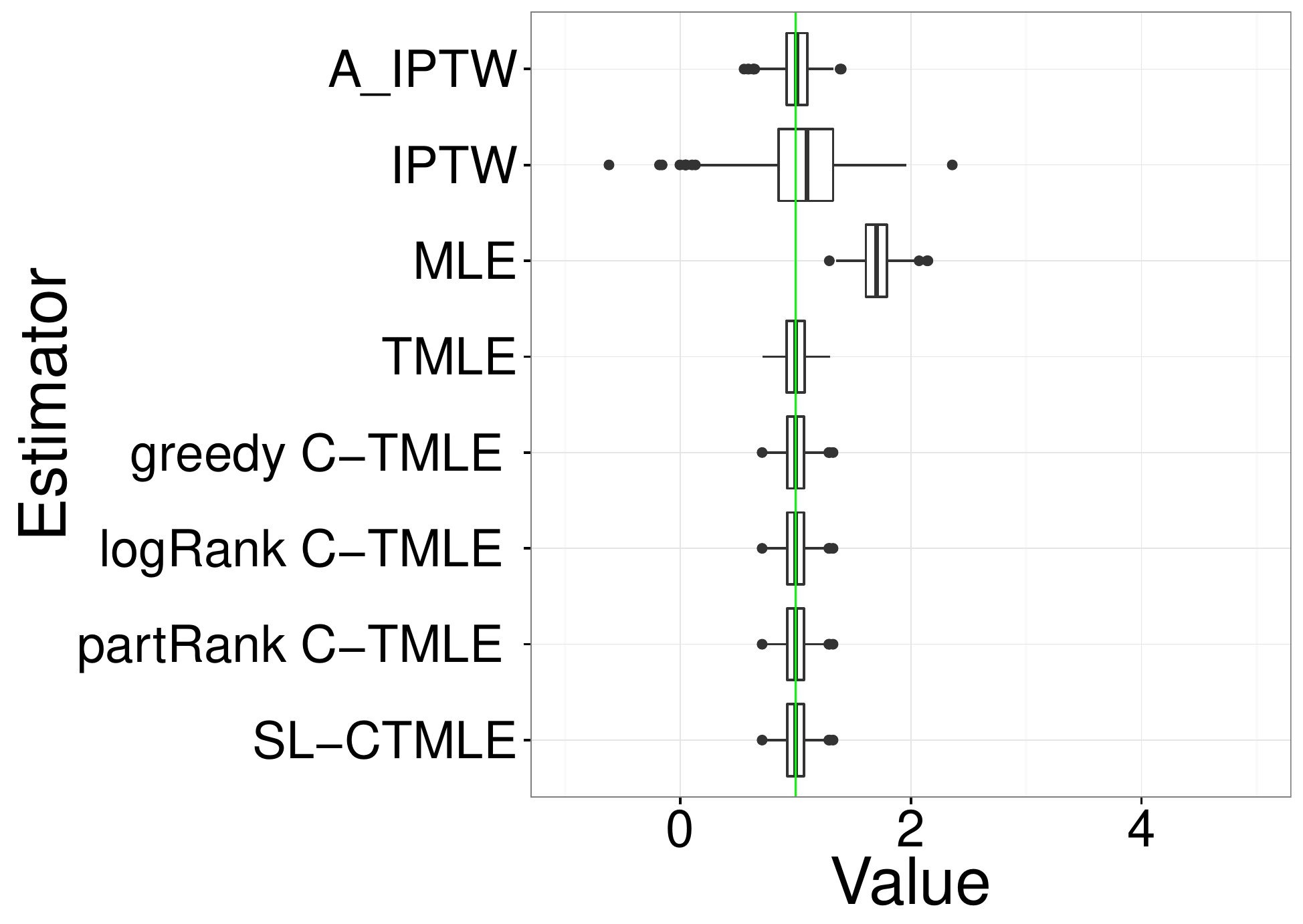}
    \caption{Mis-specified model for $\bar{Q}_{0}$.}
    \label{fig:Freedman2008-Qm}
  \end{subfigure}
  \caption{Simulation 1:  Box plot of ATE  estimates with correct/mis-specified
    models for  $\bar{Q}_{0}$.  The  green line  indicates the  true parameter
    value.}
  \label{fig:Freedman2008}
\end{figure}

Bias, variance,  and mean squared error  (MSE) for all estimators  across 1000
simulated data sets are shown in Table \ref{table:Freedman2008}.  Box plots of
the estimated ATE are shown in Fig.~\ref{fig:Freedman2008}.
When $Q_{0}$  was correctly  specified, all  models had  very small  bias.  As
Freedman and  Berk discussed,  even when  the correct PS  model is  used, near
positivity  violations can  lead to  finite  sample bias  for IPTW  estimators
\cite[see  also][]{petersen2012diagnosing}.   Scalable C-TMLEs  had
smaller bias than the other DR estimators, but the distinctions were small.

When $Q_{0}$ was not correctly  specified, the G-computation/MLE estimator was
expected to be  biased.  Interestingly, A-IPTW was more biased  than the other
DR estimators. All C-TMLE estimators have identical performance, because each
approach produced the same treatment model sequence.

\subsection{Simulation Study 2: Highly correlated covariates}
\label{subsec:sim:two}

In the  second simulation study, we  study the case that  multiple confounders
are highly  correlated with each  other. We will use the notation $W_{1:k} = (W_{1}, \ldots, W_{k})$.
The data-generating  distribution is
described as follows:
\begin{eqnarray*}
  W_1,W_2,W_3 &\stackrel{iid}{\sim}& \Bernoulli(0.5),\\
  W_4|W_{1:3} &\sim& \Bernoulli(0.2 + 0.5 \cdot W_1),\\
  W_5|W_{1:4} &\sim& \Bernoulli(0.05 + 0.3 \cdot W_1 + 0.1 \cdot W_2 + 0.05 \cdot W_3 +
             0.4 \cdot W_4),\\
  W_6|W_{1:5} &\sim& \Bernoulli(0.2 + 0.6 \cdot W_5),\\
  W_7|W_{1:6} &\sim& \Bernoulli(0.5 + 0.2 \cdot W_3),\\
  W_8|W_{1:7} &\sim& \Bernoulli(0.1 + 0.2 \cdot W_2 + 0.3 \cdot W_6 + 0.1 \cdot W_7),
\end{eqnarray*}
\begin{eqnarray*}
  P_{0}(A = 1 \mid W) &=& g_0(1 \mid W) \\
                      &=& \expit( -0.05 + 0.1 \cdot W_1 +
                          0.2 \cdot W_2 + 0.2 \cdot W_{3} \\
                      && \qquad\qquad - 0.02\cdot W_4 - 0.6 \cdot W_5 - 0.2 \cdot W_6 - 0.1 \cdot W_7),
\end{eqnarray*}
and finally, for $\epsilon \sim N(0,1)$ (independent from $A$ and $W$),
\begin{equation*}
  Y = 10+ A + W_1 +W_2 +W_4 +2 \cdot W_6 +W_7 + \epsilon.
\end{equation*}
The true ATE for this simulation study is $\psi_0 = 1$.

In this case, the true  confounders are $W_1,W_2,W_4,W_6,W_7$. Covariate $W_5$
is  most closely  related  to  $W_1$ and  $W_4$.   Covariate  $W_3$ is  mainly
associated with $W_7$.  Neither $W_3$ nor  $W_5$ is a confounder (both of them
are predictive of  treatment $A$, but do not influence  directly outcome $Y$).
Including either one of them in  the PS model should inflate the
variance~\citep{brookhart2006variable}.

As  in  Section~\ref{subsec:sim:one}, each  of  the  estimators involving  the
estimation of  $\bar{Q}_{0}$ was implemented  twice, a correctly
specified  model  to  estimate  $Q_0$, and a mis-specified  model  defined by a   linear
regression model of $Y$ on $A$ only.

\begin{table}[H]
  \centering
  \caption{Simulation study  2. Performance of the  various estimators across
    1000 simulated data sets of sample size 1000.}
   \label{table:Gruber2015}
    \scalebox{0.8}{
  \begin{tabular}{l|rrr|rrr}
    \hline & \multicolumn{3}{c}{\textbf{correct $\bar{Q}$}} &
        \multicolumn{3}{c}{\textbf{mis-specified $\bar{Q}$}} \\ & bias ($10^{-3}$) & se ($10^{-2}$) & MSE ($10^{-3}$) &
    bias ($10^{-3}$) & se ($10^{-2}$) & MSE ($10^{-3}$) \\
    \hline
    unadj & 392.9              & 12.65&170.3& 392.9 & 12.65 & 170.3\\ 
    A\_IPTW & 2.4              & 6.54 & 4.3 & 2.0  & 6.53 & 4.3 \\ 
    IPTW & 2.1                 & 7.78 & 6.0 & 2.1  & 7.78 & 6.0 \\ 
    MLE & 2.6                  & 6.52 & 4.3 & 391.2 & 12.39 & 168.4\\ 
    TMLE & 2.4                 & 6.54 & 4.3 & 2.0  & 6.53 & 4.3 \\ 
    greedy C-TMLE      & 2.6   & 6.52 & 4.3 & 11.4  & 7.01 & 5.0 \\ 
    logRank C-TMLE     & 2.5   & 6.52 & 4.3 & 6.3  & 6.72& 4.6 \\ 
    partRank C-TMLE    & 2.6   & 6.52 & 4.3 & 2.5  & 6.67 & 4.4 \\ 
    SL-C-TMLE & 2.5            & 6.52 & 4.3 & 5.2  & 6.79 & 4.6 \\ 
    \hline

  \end{tabular}
 }
\end{table}

\begin{figure}[H]
  \centering
  \begin{subfigure}[t]{0.4\textwidth}
    \includegraphics[width=\textwidth]{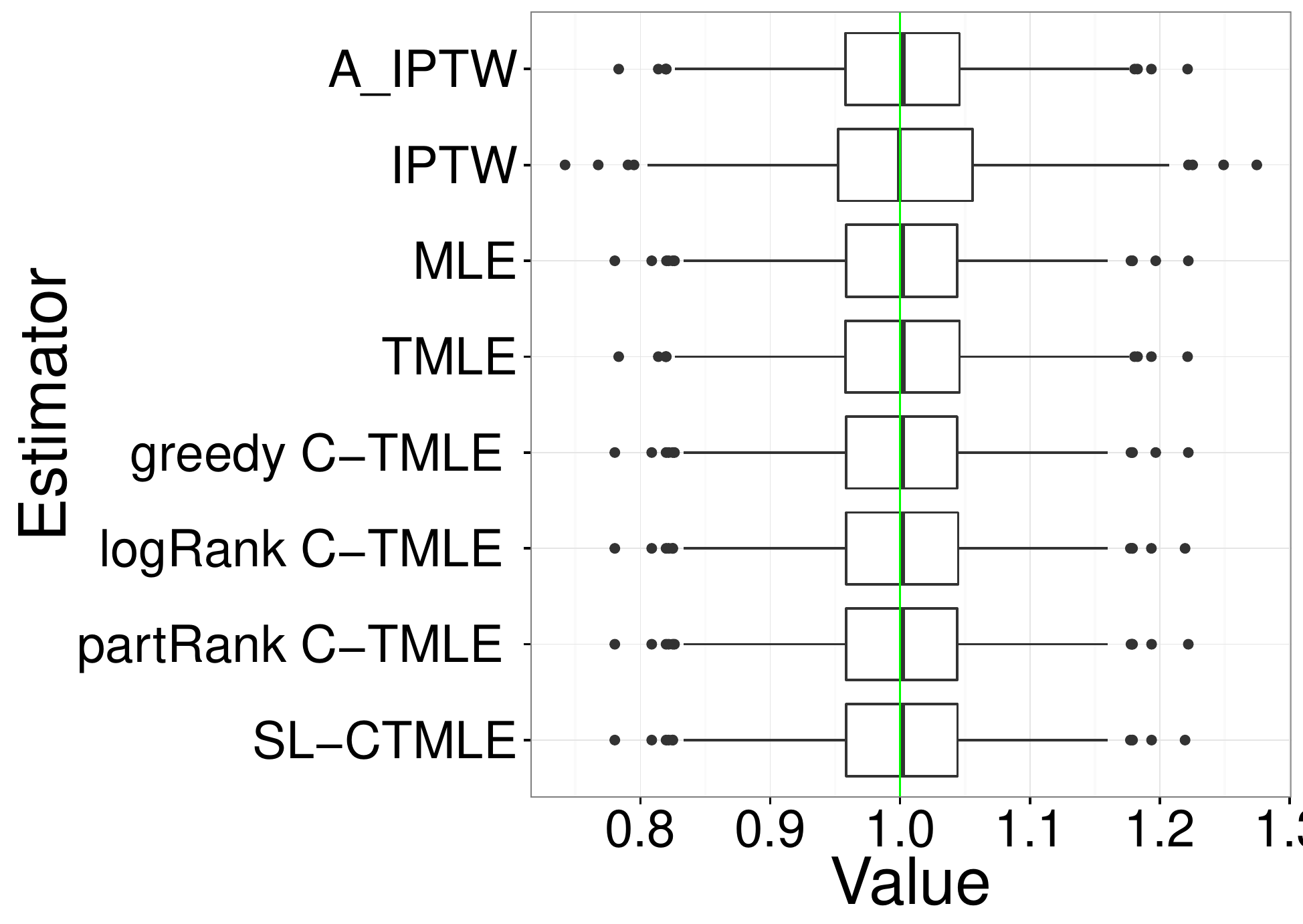}
    \caption{Well specified model for $\bar{Q}_{0}$.}
    \label{fig:Gruber2015-correct}
  \end{subfigure}
    \hspace{1cm}
  \begin{subfigure}[t]{0.4\textwidth}
    \includegraphics[width=\textwidth]{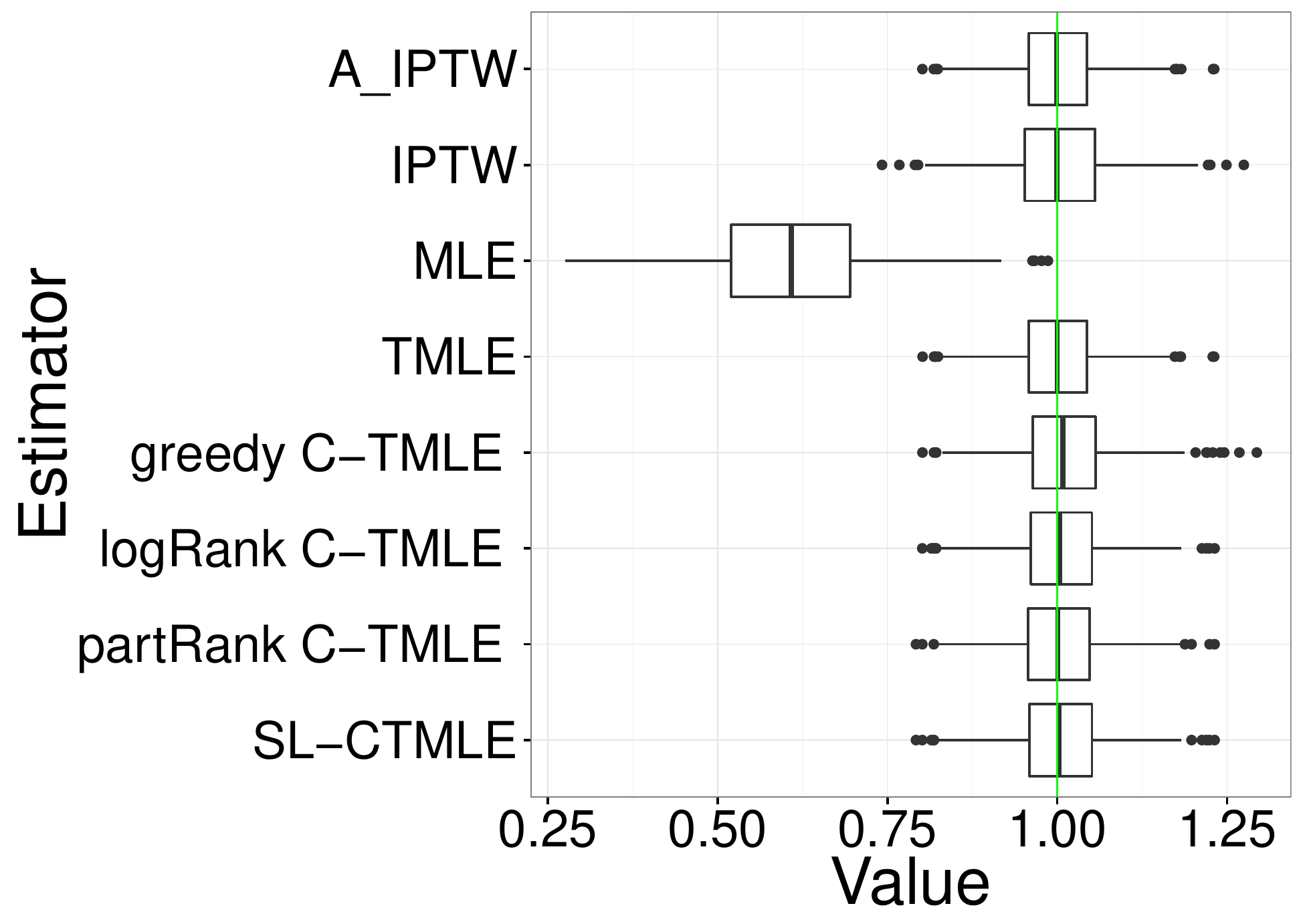}
    \caption{Mis-specified model for $\bar{Q}_{0}$. }
    \label{fig:Gruber2015-intercept}
  \end{subfigure}
  \caption{Simulation 2:  Box plot of ATE  estimates with correct/mis-specified
    models  for $\bar{Q}_{0}$.  The green  line indicates  the true  parameter
    value.}
  \label{fig:Gruber2015}
\end{figure}

Table \ref{table:Gruber2015} demonstrates and compares performance across
1000   replications.   Box   plots  of   the  estimated   ATE  are   shown  in
Fig.~\ref{fig:Gruber2015}.   When $\bar{Q}_{0}$  was correctly  specified, all
estimators except the unadjusted estimator  had small bias.  The DR estimators
had lower  MSE than  the inefficient IPTW  estimator.  When  $\bar{Q}_{0}$ was
mis-specified, the A-IPTW and IPTW estimators were less biased than the C-TMLE
estimators. The  bias of the  greedy C-TMLE was five  times larger.
However, all  DR estimators had  lower MSE than  the IPTW estimator,  with the
TMLE outperforming the others.

\subsection{Simulation Study 3: Binary outcome with instrumental variable}
\label{subsec:sim:three}

In the  third simulation, we  assess the performance of  C-TMLE in a  data set
with  positivity  violations.    We  first  generate  $W_1,   W_2,  W_3,  W_4$
independently    from   the    uniform   distribution    on   $[0,1]$,    then
$A|W \sim \Bernoulli(g_{0}(1|W))$ with
$$g_{0} (1,W) = \expit(-2 + 5W_1 + 2W_2 + 1 W_3), $$ 
and finally $Y|(A,W) \sim \Bernoulli(\bar{Q}_{0}(A,W))$ with
$$ \bar{Q}_{0}(A,W)= \expit(-3 + 2W_2+ 2 W_3 + W_4 + A).$$

As  in Sections~\ref{subsec:sim:one}  and  \ref{subsec:sim:two},  each of  the
estimators involving  the estimation  of $\bar{Q}_{0}$ was  implemented twice,
once with a correctly specified model and once with a mis-specified
 linear regression model of $Y$ on $A$ only.


\begin{table}[H]
  \centering
  \caption{Simulation study 3.  Performance  of the various estimators across
    1000 simulated data sets of sample size 10000.}
   \label{table:Ju2015}
    \scalebox{0.8}{
  \begin{tabular}{l|rrr|rrr}
    \hline & \multicolumn{3}{c}{\textbf{correct $\bar{Q}$}} &
         \multicolumn{3}{c}{\textbf{mis-specified $\bar{Q}$}} \\ & bias ($10^{-3}$) & se ($10^{-2}$) & MSE ($10^{-3}$) &
    bias ($10^{-3}$) & se ($10^{-2}$) & MSE ($10^{-3}$) \\
    \hline

     unadj & 78.1         & 3.72 & 7.5 & 78.1 & 3.72 & 7.5 \\ 
     A\_IPTW & 1.7        & 5.62 & 3.2 & 13.9 & 5.64 & 3.4 \\ 
     IPTW & 45.9          & 6.05 & 5.8 & 45.9 & 6.05 & 5.8 \\ 
     MLE & 0.7            & 4.20 & 1.8 & 76.4 & 3.61 & 7.1 \\ 
     TMLE & 1.5           & 6.28 & 3.9 &  1.3 & 6.44 & 4.1 \\ 
     greedy C-TMLE  & 0.4 & 5.39 & 2.9 & 12.2 & 5.79 & 3.5 \\ 
     logRank C-TMLE & 0.9 & 5.39 & 2.9 & 11.2 & 5.59 & 3.3 \\ 
   partRank C-TMLE  & 1.2 & 5.65 & 3.2 &  6.9 & 5.37 & 2.9 \\ 
     SL-C-TMLE & 0.3      & 5.73 & 3.3 &  7.7 & 5.46 & 3.0 \\ 
     \hline
  \end{tabular}
 }
\end{table}

\begin{figure}[H]
  \centering
  \begin{subfigure}[t]{0.4\textwidth}
    \includegraphics[width=\textwidth]{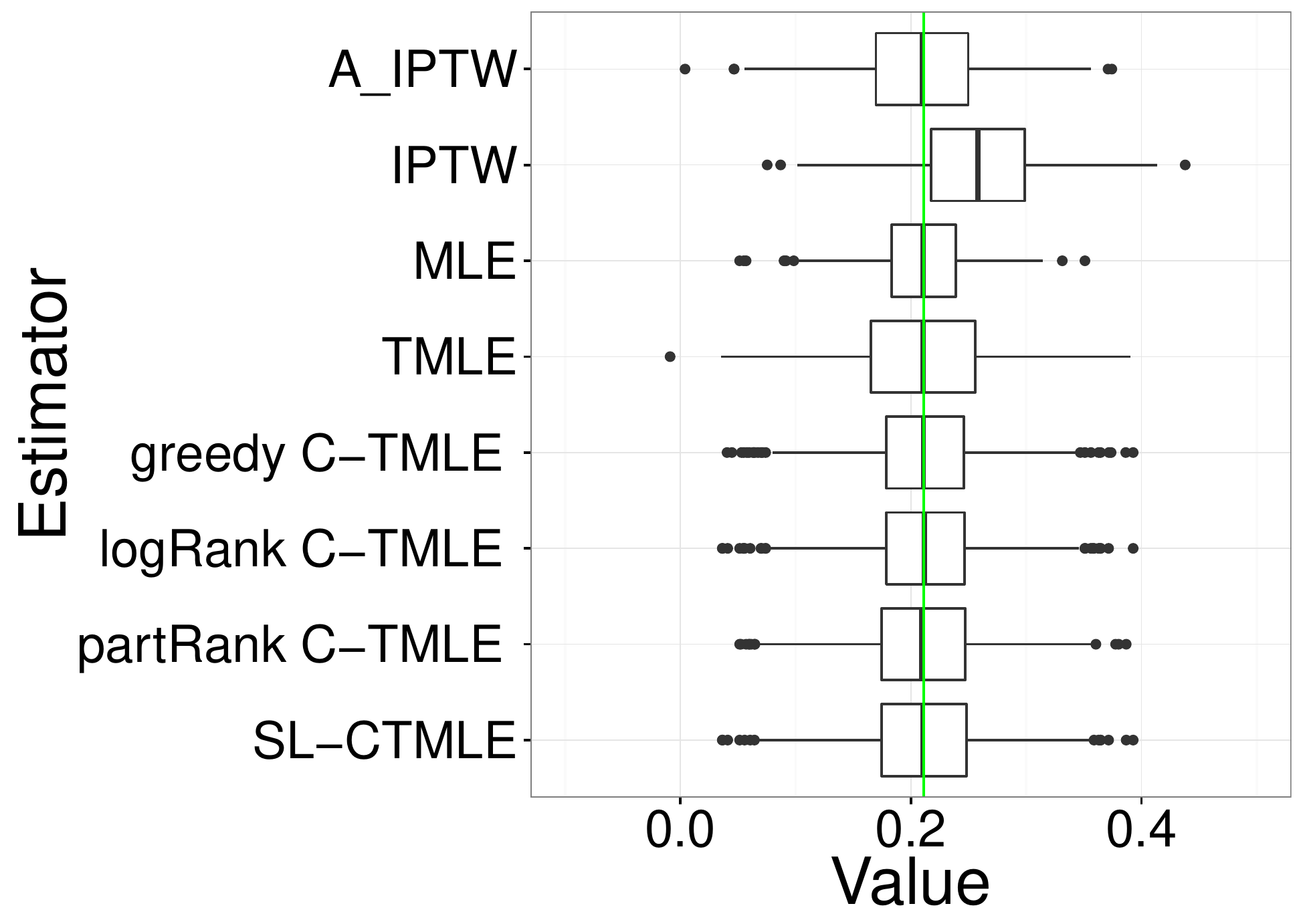}
    \caption{Well specified model for $\bar{Q}_{0}$.}
    \label{fig:Ju2015-correct}
  \end{subfigure}
    \hspace{1cm}
  \begin{subfigure}[t]{0.4\textwidth}
    \includegraphics[width=\textwidth]{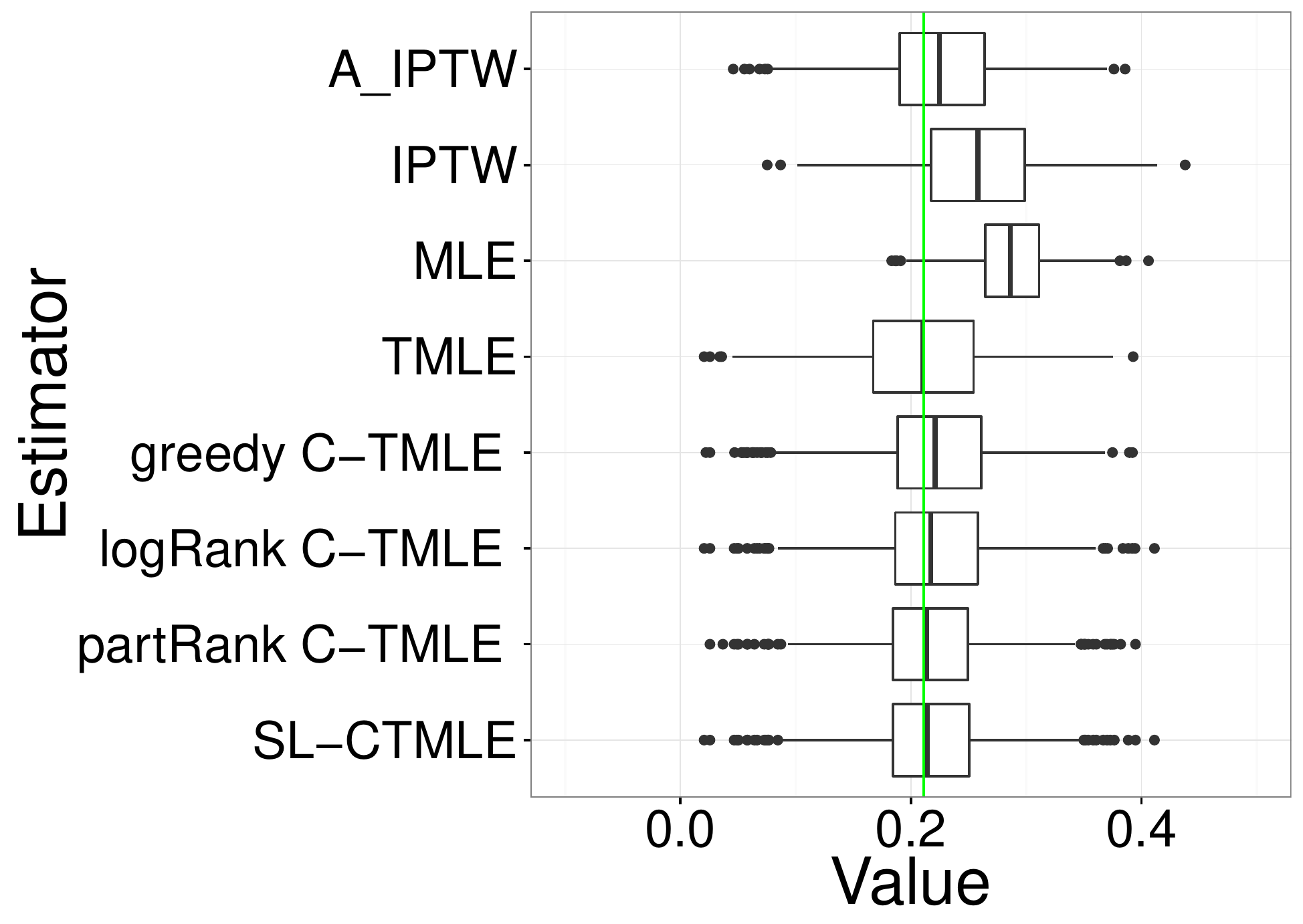}
    \caption{Mis-specified model for $\bar{Q}_{0}$.}
    \label{fig:Ju2015-intercept}
  \end{subfigure}
  \caption{Simulation 3:  Box plot of ATE  estimates with correct/mis-specified
    models  for $\bar{Q}_{0}$.  The green  line indicates  the true  parameter
    value.}
  \label{fig:Ju2015}
\end{figure}

Table  \ref{table:Ju2015}  demonstrates  the performance  of  the  estimators
across  1000  replications.   Fig.~\ref{fig:Ju2015}  shows box  plots  of  the
estimates  for the  different  methods  across 1000  simulation,  with a  well
specified or mis-specified model for $\bar{Q}_{0}$.

When  $\bar{Q}_{0}$ was  correctly specified,  the DR  estimators had  similar
bias/variance trade-offs.  Although IPTW is  a consistent estimator when $g$ is
correctly  specified,  truncation  of  the PS  $g_n$  may  have
introduced bias.  However,  without truncation it would have been extremely unstable due to
violations of the positivity assumption when instrumental variables are included in the propensity score model.

When the model for $\bar{Q}_{0}$ was  mis-specified, the MLE was equivalent to
the unadjusted estimator.  The DR methods  performed well with an MSE close to
that  observed  when  $\bar{Q}_{0}$   was  correctly  specified.   All  C-TMLEs
had similar performance.  They  out-performed the other DR methods
(namely,  A-IPTW  and  TMLE)  and the  pre-ordering  strategies  improved  the
computational  time without  loss of  precision  or accuracy  compared to  the
greedy C-TMLE algorithm.

\subsubsection*{Side note.}

Because $W_1$ is an instrumental variable that is highly predictive of the PS,
but not helpful for confounding control, we expect that including it in the PS
model  would increase  the variance  of the  estimator.  One  possible way  to
improve  the performance  of the  IPTW estimator  would be  to apply  a C-TMLE
algorithm to select covariates for fitting the PS model.  In the mis-specified
model for $\bar{Q}_{0}$ scenario, we also simulated the following procedure:
\begin{enumerate}
\item Use a greedy C-TMLE algorithm to select the covariates.
\item Use main  terms logistic regression with selected covariates  for the PS
  model.
\item Compute IPTW using the estimated PS.
\end{enumerate}

The simulated bias  for this estimator was $0.0340$, the  SE was $0.0568$, and
the MSE  was $0.0043$. Excluding the  instrumental variable from the  PS model
thus reduced bias, variance, and MSE of the IPTW estimator.

\subsection{Simulation   Study  4:   Continuous   outcome  with   instrumental
  covariate}
\label{subsec:sim:four}

In the fourth simulation, we assess  the performance of C-TMLEs in a
simulation     scheme      with     a     continuous      outcome     inspired
by~\citep{gruber2011c} (we  merely increased  the coefficient in  front of
$W_1$ to introduce  a stronger positivity violation).   We first independently
draw $W_1,  W_2, W_3,  W_4,W_5, W_6$  from the standard  normal law,  then $A$
given $W$ with
\begin{equation*}
  P_0(A = 1 \mid W) = g_{0}(1,W) = \expit(2W_1 + 0.2W_2 + -3 W_3),
\end{equation*}
and finally $Y$ given $(A,W)$ from a Gaussian law with variance 1 and mean
\begin{equation*}
  \bar{Q}_{0}(A,W) = 0.5 W_1 - 8 W_2 + 9 W_3 - 2 W_5 + A.
\end{equation*}

The initial estimator $\bar{Q}_{n}^{0}$ was built based on a linear regression
model  of  $Y$  on  $A$,  $W_1$,  and  $W_2$,  thus  partially  adjusting  for
confounding.  There  was residual  confounding due to  $W_3$.  There  was also
residual confounding  due to $W_1$  and $W_2$ within  at least one  stratum of
$A$, despite their inclusion in the initial outcome regression model.
 


\begin{table}[H]
  \centering
  \caption{Simulation study 4.  Performance  of the various estimators across
    1000 simulated data sets of sample  size 1000.  Omitted in the table, the
    performance of the unadjusted estimator  was an order of magnitude worse
    than the performance of the other estimators.}
   \label{table:sim4}
    \scalebox{0.8}{
  \begin{tabular}{l|rrr|}
    \hline & \multicolumn{3}{c}{\textbf{Mis-specified $\bar{Q}$}} \\
    & bias & se & MSE \\
         \hline
         A\_IPTW             & 4.49   & 0.84& 20.88  \\
         IPTW                & 2.97   & 0.87 & 9.60 \\
     MLE                     & 12.68  & 0.47 & 161.20  \\
     TMLE                    & 1.31   & 1.21 & 3.17 \\
     greedy C-TMLE           & 0.25  & 1.01 & 1.27 \\
     logRank C-TMLE          & 0.36  & 0.88 & 0.90 \\
   partRank C-TMLE           & 0.32  & 0.92 & 0.95  \\
     SL-C-TMLE               & 0.37  & 0.88 & 0.90 \\
     \hline
  \end{tabular}
 }
\end{table}

\begin{figure}[H]
  \centering 
    \includegraphics[width=0.6\textwidth]{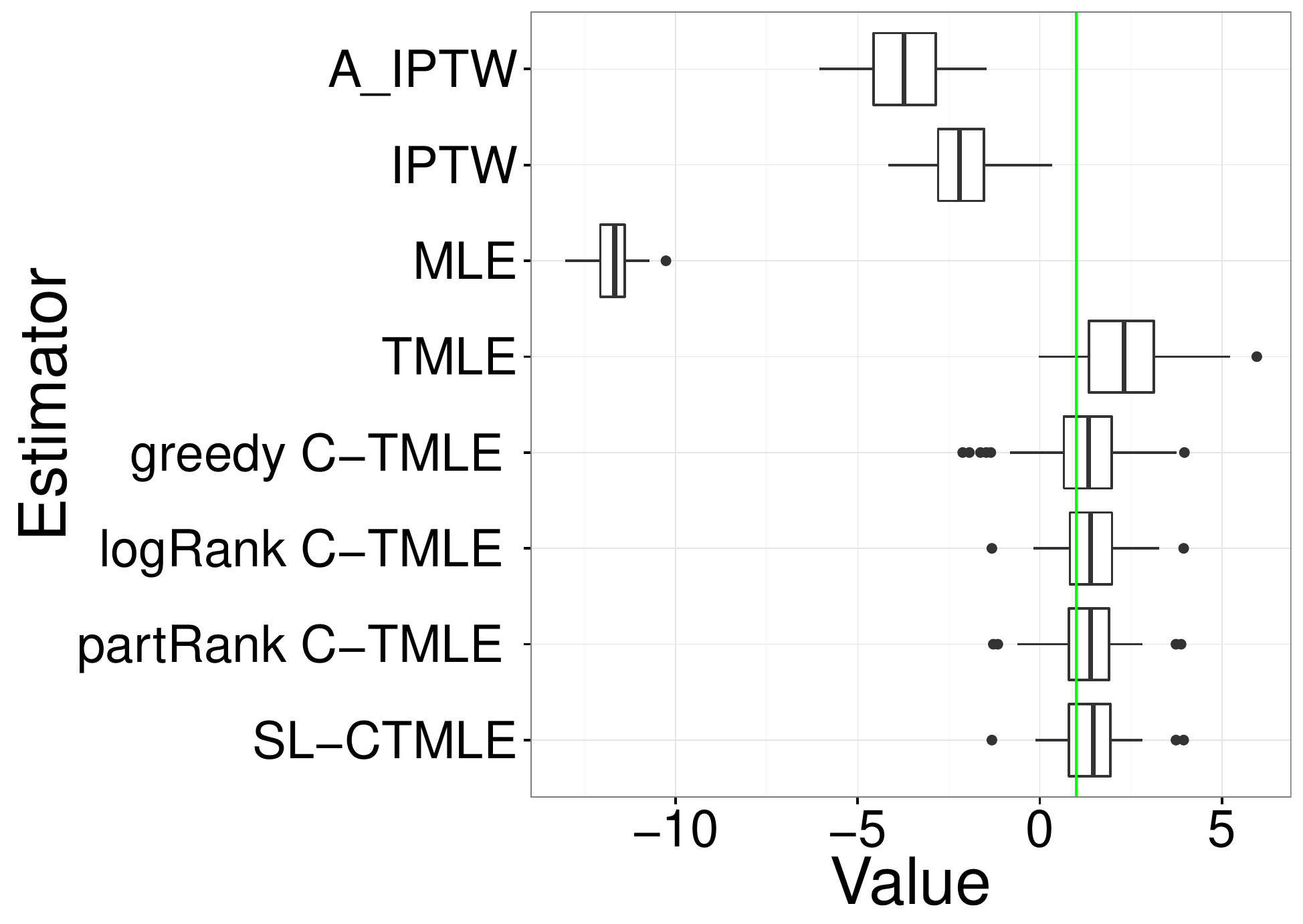}
    \caption{Simulation 4: Box  plot of ATE estimates  with mis-specified model
      for $\bar{Q}_{0}$.}
    \label{fig:sim4}
\end{figure}

Fig.~\ref{fig:sim4} reveals  that the C-TMLEs   performed much better
than TMLE  and A-IPTW estimators  in terms of  bias and standard  error.  This
illustrates that choosing  to adjust for less than the  full set of covariates
can  improve   finite sample  performance when  there are  near positivity
violations.  In  addition, Table  \ref{table:sim4} shows that  the pre-ordered
C-TMLEs out-performed the greedy C-TMLE.  Although the greedy C-TMLE estimator
had  smaller  bias,   it  had  higher  variance,  perhaps  due   to  its  more
data-adaptive ordering procedure.

\section{Simulation Study on Partially Synthetic Data}
\label{subsec:sim:five}

The aim of this  section is to compare TMLE and all  C-TMLEs  using a
large   simulated   data    set   that   mimics   a    real-world   data   set.
Section~\ref{subsec:sim:real}  starts the  description of  the data-generating
scheme and  resulting large data set.   Section~\ref{subsec:hdPs} presents the
High-Dimensional Propensity Score  (hdPS) method used to  reduce the dimension
of the data set.  Section~\ref{subsec:sim:real:cont} completes the description
of the data-generating scheme and  specifies how $\bar{Q}_{0}$ and $g_{0}$ are
estimated.   Section~\ref{subsec:results:real} summarizes  the results  of the
simulation study.

\subsection{Data-generating scheme}
\label{subsec:sim:real}

The  simulation  scheme relies  on  the  Nonsteroidal anti-inflammatory  drugs
(NSAID)         data         set         presented         and         studied
in~\citep{schneeweiss2009high,rassen2012using}.   Its $n=49,653$  observations
were  sampled from  a  population of  patients  aged 65  years  and older,  and
enrolled  in  both Medicare  and  the  Pennsylvania Pharmaceutical  Assistance
Contract for the Elderly (PACE) programs  between 1995 and 2002. Each observed
data structure consists of a triplet  $(W,A,Y)$ where $W$ is decomposed in two
parts:  a vector  of 22  baseline  covariates and  a highly  sparse vector  of
$C=9,470$ unique  claim codes.   In the  latter, each  entry is  a nonnegative
integer indicating how many times  (mostly zero) a certain procedure (uniquely
identified  among $C=9,470$  by  its claim  code) has  been  undergone by  the
corresponding  patient.   The claim  codes  were  manually clustered  into  eight
categories: ambulatory  diagnoses, ambulatory procedures,  hospital diagnoses,
hospital procedures,  nursing home  diagnoses, physician  diagnoses, physician
procedures  and  prescription drugs.   The  binary  indicator $A$  stands  for
exposure to a selective COX-2 inhibitor  or a comparison drug (a non-selective
NSAID).  Finally,  the binary outcome  $Y$ indicates  whether or not  either a
hospitalization for severe gastrointestinal hemorrhage or peptic ulcer disease
complications including perforation in GI patients occurred.

The        simulated       data        set       was        generated       as
in~\citep{gadbury2008evaluating,franklin2014plasmode}.   It took  the form  of
$n=49,653$     data    structures     $(W_{i},     A_{i},    Y_{i})$     where
$\{(W_{i}, A_{i}) : 1 \leq i \leq  n\}$ was extracted from the above real data
set and where $\{Y_{i} : 1 \leq i \leq  n\}$ was simulated by us in such a way
that, for each $1 \leq i \leq n$, the random sampling of $Y_{i}$ depended only
on  the  corresponding $(W_{i},  A_{i})$.   As  argued in  the  aforementioned
articles, this approach  preserves the covariance structure  of the covariates
and complexity of  the true treatment assignment mechanism,  while allowing  the true value of the ATE
parameter to be known, and preserving  control over the degree of
confounding.

\subsection{High-Dimensional   Propensity   Score  Method   For   Dimension
  Reduction}
\label{subsec:hdPs}

The  simulated data  set was  large, both  in number  of observations  and the
number of covariates. In this framework, directly applying any version of
C-TMLE  algorithms  would  not  be  the  best  course  of  action  First,  the
computational  time would  be unreasonably  long due  to the  large number  of
covariates.   Second, the  resulting estimators  would  be plagued  by  high
variance due to the low signal-to-noise ratio in the claim data.

This motivated us to apply the High-Dimensional Propensity Score (hdPS) method
for  dimension reduction  prior to  applying  the TMLE  and C-TMLE  algorithms. 

Introduced  in~\citep{schneeweiss2009high}), the  hdPS method
was proposed to reduce the dimension in large electronic healthcare databases.
It    is     is    increasingly    used    in     studies    involving    such
databases~\citep{rassen2012using,patorno2014studies,franklin2015regularized,toh2011confounding,kumamaru2016comparison,ju2016propensity}). 

The hdPS method  essentially consists of two main steps:  {\em (i)} generating
so  called hdPS  covariates  from  the claims  data  (which  can increase  the
dimension) then {\em (ii)} screening  the enlarged collection of covariates to
select a small proportion of  them (which dramatically reduces the dimension).
Specifically, the method unfolds as follows~\citep{schneeweiss2009high}:
\begin{description}[%
  before={\setcounter{descriptcount}{0}},%
  font=\bfseries\stepcounter{descriptcount}\thedescriptcount.~]
\item[Cluster  by Resource.]   Cluster  the  data by  resource  in ${\cal  C}$
  clusters.

  In the  current example, we  derived ${\cal C}=8$ clusters  corresponding to
  the  following  categories:  ambulatory  diagnoses,  ambulatory  procedures,
  hospital diagnoses,  hospital procedures, nursing home  diagnoses, physician
  diagnoses, physician procedures and prescription drugs.
  See~\citep{schneeweiss2009high,patorno2014studies} for other examples.\\
\item[Identify Candidate Claim Codes.]  For  each cluster separately, for each
  claim code $c$ within the  cluster, compute the empirical proportion $Pr(c)$
  of  positive entries,  then sort  the claim  codes by  decreasing values  of
  $\min(Pr(c), 1-Pr(c))$.  Finally,  select only the top $J$  claim codes.  We
  thus go from $C$ claim codes to $J\times {\cal C}$ claim codes.  

  As explained below, we chose $J=50$ so the dimension of the claims data went
  from $9,470$ to 400.\\
\item[Assess Recurrence of Claim Codes.]  For each selected claim code $c$ and
  each patient $1 \leq i \leq n$, {\em replace} the corresponding $c_{i}$ with
  three binary  covariates called ``hdPS covariates'':  $c_{i}^{(1)}$ equal to
  one if and only if (iff) $c_{i}$ is positive; $c_{i}^{(2)}$ equal to one iff
  $c_{i}$  is  larger  than the  median  of  $\{c_{i}  :  1\leq i  \leq  n\}$;
  $c_{i}^{(3)}$ equal to  one iff $c_{i}$ is larger than  the 75\%-quantile of
  $\{c_{i}  : 1\leq  i \leq  n\}$.  This  inflates the  number of  claim codes
  related covariates by a factor 3.

  As explained below, the dimension of the claims data thus went from 400 to
  $1,200$.\\
\item[Select Among the hdPS Covariates.]   For each hdPS covariate, estimate a
  measure of  its potential confounding  impact, then sort them  by decreasing
  values of  the estimates of the  measure.  Finally, select only  the top $K$
  hdPS covariates.

  For instance, one can  rely on the following estimate of  the measure of the
  potential  confounding   impact  introduced  in~\citep{bross54}:   for  hdPS
  covariate $c^{\ell}$
  \begin{equation}
    \label{eq:Bross}
    \frac{\pi_{n}^{\ell}(1) (r_{n}^{\ell} - 1) +
      1}{\pi_{n}^{\ell}(0) (r_{n}^{\ell} - 1) +
      1}
  \end{equation}
  where
  \begin{eqnarray*}
    \pi_{n}^{\ell} (a)
    &=&    \frac{\sum_{i=1}^{n}   \1\{c_{i}^{\ell}=1,a_{i}=a\}}{\sum_{i=1}^{n}
        \1\{a_{i}=a\}} \quad (a=0,1) \quad \text{and}\\
    r_{n}^{\ell}
    &=& \frac{p_{n}(1)}{p_{n}(0)} \quad \text{with} \quad 
        p_{n}(c)
        = \frac{\sum_{i=1}^{n}   \1\{y_{i}=1,c_{i}^{\ell}=c\}}{\sum_{i=1}^{n}
        \1\{c_{i}^{\ell}=c\}} \quad (c=0,1).
  \end{eqnarray*}
  A  rationale for  this choice  can be  found in~\citep{schneeweiss2009high},
  where     $r_{n}^{\ell}$    in     \eqref{eq:Bross}    is     replaced    by
  $\max(r_{n}^{\ell}, 1/r_{n}^{\ell})$.

  
  As explained below we chose $K=100$. As a result, the dimension of the claims
  data was reduced to 100 from $9,470$.
\end{description}

\subsection{Data-generating scheme (continued) and estimating procedures}
\label{subsec:sim:real:cont}

Let us  resume here  the presentation  of the  simulation scheme  initiated in
Section~\ref{subsec:sim:real}. Recall  that the  simulated data set  writes as
$\{(W_{i},    A_{i},    Y_{i})    :    1    \leq    i    \leq    n\}$    where
$\{W_{i}  : 1  \leq i  \leq  n\}$ is  the by  product  of the  hdPS method  of
Section~\ref{subsec:hdPs}      with      $J=50$      and      $K=100$      and
$\{A_{i} : 1  \leq i \leq n\}$  is the original vector of  exposures.  It only
remains to present how $\{Y_{i} : 1 \leq i \leq n\}$ was generated.

First, we  arbitrarily chose a subset  $W^\prime$ of $W$, that  consists of 10
baseline  covariates ({\em  congestive heart  failure}, {\em  previous use  of
  warfarin}, {\em number of generic drugs  in last year}, {\em previous use of
  oral  steroids},  {\em  rheumatoid  arthritis}, {\em  age  in  years},  {\em
  osteoarthritis}, {\em number  of doctor visits in last  year}, {\em calendar
  year}) and 5 hdPS covariates.  Second, we arbitrarily defined a parameter
\begin{multline*}
  \beta = (1.280, -1.727, 1.690, 0.503, 2.528, 0.549, 0.238, -1.048, 1.294,\\
  0.825, -0.055, -0.784, -0.733, -0.215, -0.334)^{\top}.
\end{multline*}
Finally,   $Y_{1},   \ldots,   Y_{n}$   were   independently   sampled   given
$\{(W_{i},  A_{i}) :  1 \leq  i \leq  n\}$ from  Bernoulli distributions  with
parameters $q_{1}, \ldots, q_{n}$ where, for each $1 \leq i \leq n$,
\begin{equation*}
  q_{i} = \expit \left(\beta^{\top} W_{i}' + A_{i}\right).
\end{equation*}
The resulting true value of the ATE is $\psi_0 = 0.21156$.\\

The estimation  of the conditional  expectation $\bar{Q}_{0}$ was  carried out
based on  two logistic regression  models.  The  first one was  well specified
whereas the second one was mis-specified, due to the omission of the five hdPS
covariates.

For the TMLE algorithm, the estimation of the PS $g_{0}$ was carried out based
on a  single, main terms  logistic regression model  including all of  the 122
covariates. For  the C-TMLE algorithms,  main terms logistic  regression model
were also fitted at each step.  An early stopping rule was implemented to save
computational   time.    Specifically,   if  the   cross-validated   loss   of
$\bar{Q}_{n,k}^{*}$   is   smaller   than  the   cross-validated   losses   of
$\bar{Q}_{n,k+1}^{*},  \ldots, \bar{Q}_{n,k+10}^{*}$,  then  the procedure  is
stopped and outputs the TMLE estimator corresponding to $\bar{Q}_{n,k}^{*}$.

The scalable  SL-C-TMLE library included  the two scalable  pre-ordered C-TMLE
algorithms and excluded the greedy C-TMLE algorithm.

\subsection{Results}
\label{subsec:results:real}


Table~\ref{table:sim_data}  reports  the  point estimates  for  $\psi_{0}$  as
derived by all the considered methods.  It also reports the 95\% CIs of the form  $[\psi_{n} \pm 1.96  \sigma_{n}/\sqrt{n}]$, where
$\sigma_{n}^{2}   =   n^{-1}   \sum_{i=1}^{n}   D^{*}   (\bar{Q}_{n},   g_{n})
(O_{i})^{2}$ estimates  the variance of  the efficient influence curve  at the
couple $(\bar{Q}_{n},  g_{n})$ yielding  $\psi_{n}$.  We refer  the interested
reader to~\citep[][Appendix~A]{van2011targeted} for details on influence curve
based  inference.   All  the  CIs  contained the  true  value  of  $\psi_{0}$.
Table~\ref{table:sim_data} also reports processing times (in seconds).

\begin{table}[H]
  \caption{Point estimates  and 95\% CIs for  TMLE and
    C-TMLE estimators}
  \label{table:sim_data}
  \resizebox{\columnwidth}{!}{
    \begin{tabular}{|l|l|l|l|r|}
      \hline
      ~ & Model for $\bar{Q}_{n}^{0}$ & estimate & CI & Processing time\\ \hline
      TMLE & Well specified & 0.202 & (0.193, 0.212) & 0.6s \\
      ~                                   & Mis-specified & 0.203         & (0.193, 0.213)     &  0.6s      \\ \hline
      C-TMLE,                       &  Well specified     & 0.205         & (0.196, 0.213)    & 618.7s     \\
      greedy                                   & Mis-specified & 0.214         & (0.205, 0.223)     & 1101.2s    \\ \hline
      C-TMLE,  & Well specified & 0.205 & (0.196, 0.213) & 57.4s \\
      logistic ordering                                   & Mis-specified & 0.211         & (0.202, 0.219)    & 125.6s      \\ \hline
      C-TMLE,  & Well specified & 0.205 & (0.197, 0.213) & 22.5s \\
      partial correlation ordering                                   & Mis-specified & 0.211         & (0.202, 0.219)     & 149.0s      \\ \hline
      SL-C-TMLE        & Well specified      & 0.205         & (0.197, 0.213)     & 69.8s       \\
      ~                                   & Mis-specified & 0.211         & (0.202, 0.219)     & 264.3s      \\ \hline
    \end{tabular}
}
\end{table}

The point estimates  and CIs were similar across all  C-TMLEs.  When
the model for  $\bar{Q}_{0}$ was correctly specified,  the SL-C-TMLE 
selected the partial  correlation ordering.  When the  model for $\bar{Q}_{0}$
was  mis-specified, it  selected the  logistic ordering.   In both  cases, the
estimator with smaller bias  was data-adaptively selected. In addition,
as all  the candidates in its  library were scalable, the  SL-C-TMLE algorithm
was  also scalable,  and ran  much faster  than the  greedy C-TMLE  algorithm.
Computational time for the scalable C-TMLE algorithms was approximately 1/10th
of  the computational  time  of  the greedy  C-TMLE  algorithm.

\section{Discussion}
\label{sec:discussion}

Robust inference  of a  low-dimensional parameter  in a  large semi-parametric
model traditionally relies  on external estimators  of infinite-dimensional features  of the
distribution of the data.  Typically, only  one of the latter is optimized for
the  sake of  constructing a  well  behaved estimator  of the  low-dimensional
parameter  of  interest.   For  instance, the  targeted  minimum  loss  (TMLE)
estimator of the  average treatment effect (ATE)~\eqref{eq:TMLE}  relies on an
external estimator $\bar{Q}_{n}^{0}$ of  the conditional mean $\bar{Q}_{0}$ of
the outcome given binary treatment and baseline covariates, and on an external
estimator $g_{n}$ of the propensity  score $g_{0}$.  Only $\bar{Q}_{n}^{0}$ is
optimized/updated into $\bar{Q}_{n}^{*}$  based on $g_{n}$ in such  a way that
the  resulting substitution  estimator  of the  ATE can  be  used, under  mild
assumptions, to  derive a narrow  confidence interval with a  given asymptotic
level.

There is room  for optimization in the  estimation of $g_{0}$ for  the sake of
achieving a better bias-variance trade-off in  the estimation of the ATE. This
is the  core idea  driving the  general C-TMLE template.   It uses  a targeted
penalized loss function  to make smart choices in  determining which variables
to adjust for in the estimation  of $g_{0}$, only adjusting for variables that
have not  been fully  exploited in the  construction of  $\bar{Q}_{n}^{0}$, as
revealed in the course of a data-driven sequential procedure.


The original instantiation of the general  C-TMLE template  was presented as
a greedy  forward stepwise algorithm. It  does not scale well  when the number
$p$ of covariates  increases drastically.  This motivated  the introduction of
novel instantiations of  the C-TMLE general template where  the covariates are
pre-ordered.   Their time  complexity is  $\mathcal{O}(p)$ as  opposed to  the
original $\mathcal{O}(p^2)$,  a remarkable gain.  We  proposed two pre-ordering
strategies  and  suggested   a  rule  of  thumb  to   develop  other  meaningful
strategies. Because it is usually unclear a priori which pre-ordering strategy
to  choose,  we  also  introduced   a  SL-C-TMLE  algorithm  that  enables  the
data-driven choice of the better pre-ordering  given the problem at hand.  Its
time complexity is $\mathcal{O}(p)$ as well.

The C-TMLE algorithms used in our data analyses have been implemented in Julia and 
are publicly available at \url{https://lendle.github.io/TargetedLearning.jl/}. We undertook five  simulation studies.  Four  of them
involved fully synthetic data.  The  last one involves partially synthetic data
based  on a  real  electronic  health database  and  the  implementation of  a
high-dimensional propensity score (hdPS) method for dimension reduction widely
used for  the statistical analysis  of claim codes  data. In the  appendix, we
compare the  computational times  of variants of  C-TMLE algorithms.   We also
showcase  the  use  of  C-TMLE  algorithms on  three  real  electronic  health
database.  In all  analyses involving electronic health  databases, the greedy
C-TMLE algorithm was  unacceptably slow.  Judging from  the simulation studies,
our scalable C-TMLE algorithms work well, and so does the SL-C-TMLE algorithm.

This article focused on ATE with a  binary treatment.  In future work, we will
adapt the theory and practice of scalable C-TMLE algorithms for the estimation
of the  ATE with multi-level  or continuous  treatment by employing  a working
marginal structural  model. We will  also extend  the analysis to  address the
estimation of other classical parameters of interest.

\bibliographystyle{DeGruyter} 
\bibliography{references}

\newpage
\appendix

\begin{center}
  {\huge\bf Appendix}
\end{center}

We gather  here some additional material.   Appendix~\ref{sec:package} provides
notes on a Julia software package that  implements all the proposed C-TMLE algorithms.
Appendix~\ref{sec:time} presents and compares  the empirical processing time of
C-TMLE  algorithms  for  different  sample   sizes  and  number  of  candidate
estimators of the nuisance parameter.  Appendix~\ref{sec:analyses} compares the
performance of  the new  C-TMLEs  with  standard TMLE on
three real data sets.

\section{C-TMLE Software}
\label{sec:package}

A flexible Julia software package implementing all C-TMLE algorithms described in this
article             is            publicly             available            at
\url{https://lendle.github.io/TargetedLearning.jl/}.   The   website  contains
detailed  documentation  and  a  tutorial  for researchers  who  do  not  have
experience  with  Julia.  

In    addition    to   the    two    pre-ordering    methods   described    in
Section~\ref{sec:scalableCTMLE}, the software accepts any user-defined ranking
algorithm.   The  software  also  offers   several  options  to  decrease  the
computational    time    of    the   scalable    C-TMLE    algorithms.     The
\texttt{"Pre-Ordered"}  search strategy  has an  optional argument  \texttt{k}
which defaults to 1.  At each  step, the next $k$ available ordered covariates
are added  to the model used  to estimate $g_0$.   Large $k$ can speed  up the
procedure when there  are many covariates. However, this approach  is prone to
over-fitting, and may miss the optimal solution.

An  early stopping  criteria that  avoids computing  and cross-validating  the
complete  model  containing  all  $p$ covariates  can  also  save  unnecessary
computations.  A  \texttt{"patience"} argument accelerates the  training phase
by  setting the  number of  steps  to carry  out  after having  found a  local
optimum.       To     prepare      Section~\ref{subsec:sim:real},     argument
\texttt{"patience"} was set to 10. More details are provided in that section.

\section{Time Complexity}
\label{sec:time}

\begin{figure}[htbp]
  \centering
  \begin{subfigure}[t]{0.4\textwidth}
    \includegraphics[width=\textwidth]{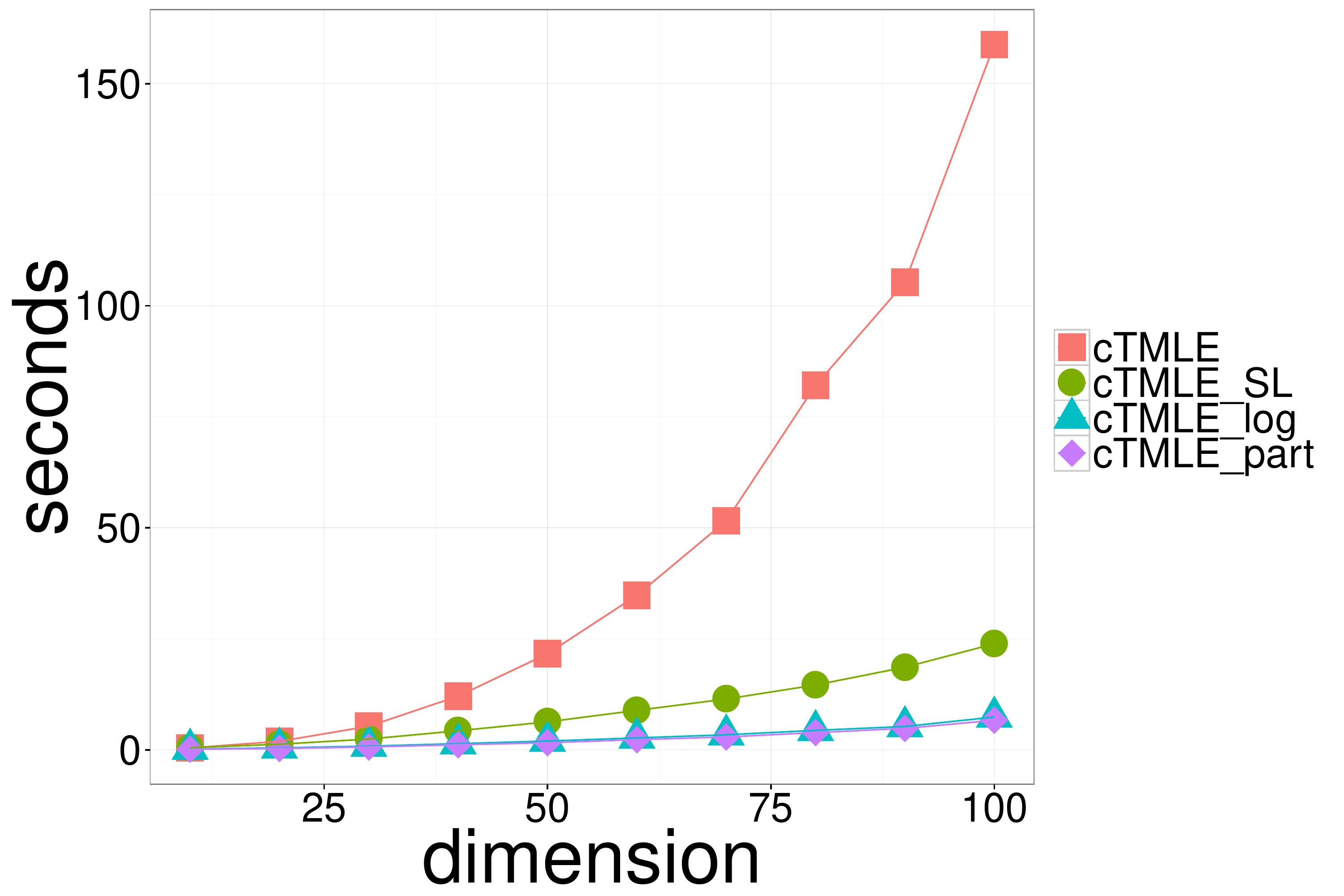}
    \caption{Median  computational  time  (across  10  replications  for  each
      point), with $n=1,000$ fixed and $p$ varying. }
    \label{fig:ptime}
  \end{subfigure}
    \hspace{1cm}
  \begin{subfigure}[t]{0.4\textwidth}
    \includegraphics[width=\textwidth]{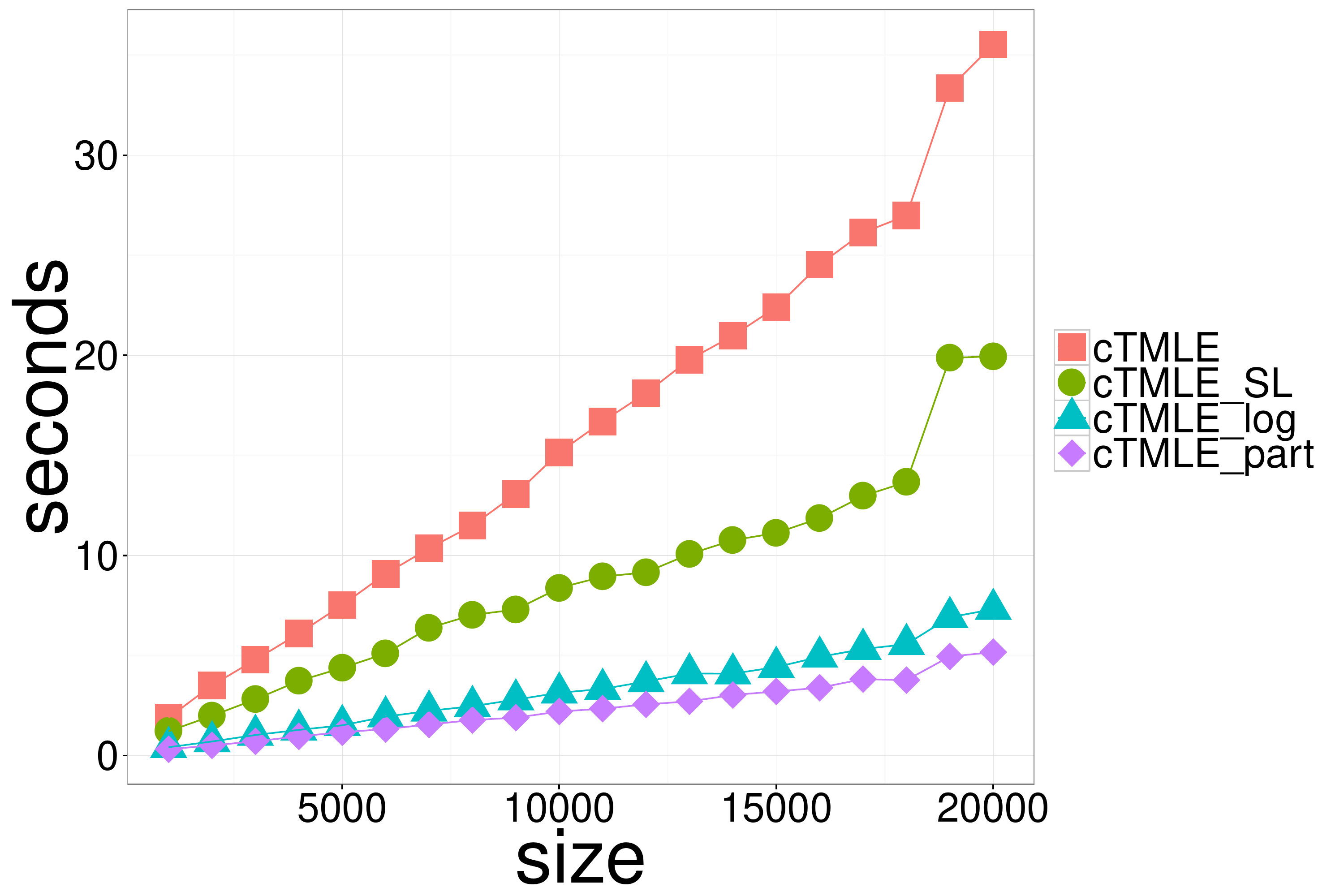}
    \caption{Median  computational  time  (across  10  replications  for  each
      point), with varying $n$ and fixed $p=20$.}
    \label{fig:ntime}
  \end{subfigure}
  \caption{Computational times of the C-TMLE algorithms with greedy search and
    pre-ordering. 
  }
  \label{fig:ctmle_time}
\end{figure}

We study  here the  computational time of  the pre-ordered  C-TMLE algorithms.
The computational  time of each algorithm  depends on the sample  size $n$ and
number of covariates $p$.  First, we set $n = 1,000$ and varied $p$ between 10
and 100  by steps of 10.   Second, we varied  $n$ from $1,000$ to  $20,000$ by
steps of $1,000$  and set $p =  20$.  For each $(n,p)$ pair,  the analysis was
replicated  ten times  independently, and  the median  computational time  was
reported.   In  every  data  set,   all  the  random  variables  are  mutually
independent.    The  results   are   shown   in  Figures~\ref{fig:ptime}   and
\ref{fig:ntime}.

Figure~\ref{fig:ptime} is in  line with the theory: the  computational time of
the forward  stepwise C-TMLE  is $\mathcal{O}(p^2)$ whereas  the computational
times of  the pre-ordered C-TMLE  algorithms are $\mathcal{O}(p)$.   Note that
the pre-ordered C-TMLEs  are indeed scalable. When $n=1,000$  and $p=100$, all
the scalable C-TMLE algorithms ran in less than 30 seconds.

Figure~\ref{fig:ntime}  reveals that  the pre-ordered  C-TMLE algorithms  are
much  faster  in practice  than  the  greedy  C-TMLE  algorithm, even  if  all
computational times are $\mathcal{O}(n)$ in that framework with fixed $p$.

\section{Real Data Analyses}
\label{sec:analyses}

This  section presents  the application  of variants  of the  TMLE and  C-TMLE
algorithms for  the analysis of three  real data sets.  Our  objectives are to
showcase their use  and to illustrate the consistency of  the results provided
by the  scalable and greedy C-TMLE  estimators.  We thus do  not implement the
competing unadjusted,  G-computation/MLE, IPTW and A-IPTW  estimators (see the
beginning of Section~\ref{sec:sim}).

In Sections~\ref{sec:sim} and \ref{subsec:sim:five}, we knew the true value of
the ATE. This is not the case here.

\subsection{Real data sets and estimating procedures}

We compared the performance of variants  of TMLE and C-TMLE algorithms across
three  observational  data  sets.    Here  are  brief  descriptions,  borrowed
from~\cite{schneeweiss2009high,ju2016propensity}.

\begin{description}
\item[NSAID Data Set.]  Refer to Section~\ref{subsec:sim:real} for its
  description.\\
\item[Novel Oral Anticoagulant (NOAC) Data Set.]  The NOAC data were collected
  between October, 2009 and December, 2012 by United Healthcare.  The data set
  tracked a cohort of  new users of oral anticoagulants for use  in a study of
  the comparative safety  and effectiveness of these agents.   The exposure is
  either ``warfarin'' or ``dabigatran''.  The binary outcome indicates whether
  or not  a patient had a  stroke during the  180 days after initiation  of an
  anticoagulant.

  The data set includes $n=18,447$ observations, $p=60$ baseline covariates and
  $C=23,531$ unique  claim codes.  The  claim codes are manually  clustered in
  four  categories:  inpatient   diagnoses,  outpatient  diagnoses,  inpatient
  procedures and outpatient procedures.\\
\item[Vytorin  Data Set.]   The Vytorin  data included  all United  Healthcare
  patients who initiated either treatment between January 1, 2003 and December
  31, 2012, with age over 65 on day of entry into cohort. The data set tracked
  a cohort  of new users of  Vytorin and high-intensity statin  therapies. The
  exposure is  either ``Vytorin'' or ``high-intensity  statin''.  The outcomes
  indicates  whether  or not  any  of  the events  ``myocardial  infarction'',
  ``stroke'' and ``death'' occurred.
 
  The data  set includes $n=148,327$ observations,  $p=67$ baseline covariates
  and $C=15,010$ unique  claim codes.  The claim codes  are manually clustered
  in five  categories: ambulatory  diagnoses, ambulatory  procedures, hospital
  diagnoses, hospital procedures, and prescription drugs.
\end{description}

Each data set is given by $\{(W_{i}, A_{i},  Y_{i}) : 1 \leq i  \leq n\}$ where
$\{W_{i}  : 1  \leq i  \leq  n\}$ is  the by  product  of the  hdPS method  of
Section~\ref{subsec:hdPs}      with      $J=100$     and      $K=200$      and
$\{(A_{i}, Y_{i}):  1 \leq i  \leq n\}$ is  the original collection  of paired
exposures and outcomes.

The estimations  of the  conditional expectation $\bar{Q}_{0}$  and of  the PS
$g_{0}$ were  carried out  based on logistic  regression models.   Both models
used either the baseline covariates only  or the baseline covariates {\em and}
the additional hdPS covariates.

To save  computational time, the  C-TMLE algorithms  relied on the  same early
stopping rule  described in Section~\ref{subsec:sim:real:cont}.   The scalable
SL-C-TMLE library included the two  scalable pre-ordered C-TMLE algorithms and
excluded the greedy C-TMLE algorithm.

\subsection{Results on the NSAID data set}
\label{subsec:NSAID}

\begin{figure}[htpb]
  \centering
  \includegraphics[width = 4in]{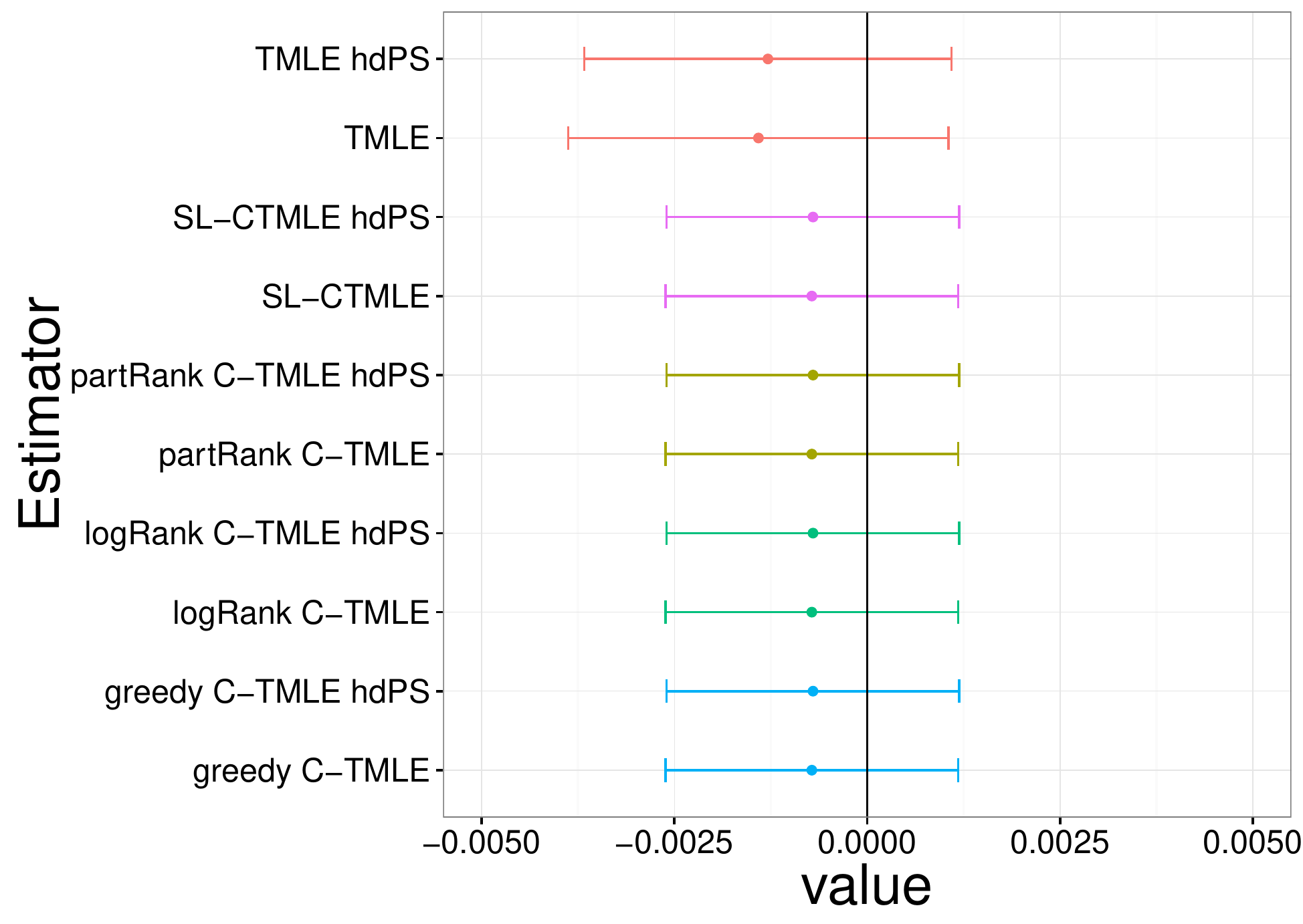}
  \caption{Point estimates and 95\% CIs   yielded by the
    different TMLE and C-TMLE estimators built on the NSAID data set. }
  \label{fig:nsaid-ctmle}
\end{figure}

Figure~\ref{fig:nsaid-ctmle} shows  the point estimates and  95\% CIs yielded  by the different TMLE and C-TMLE  estimators built from
the  NSAID data  set.

The various  C-TMLE estimators exhibit  similar results, with  slightly larger
point estimates and narrower CIs compared  to the TMLE estimators. All the CIs
contain zero.


\subsection{Results on the NOAC Data Set}
\label{subsec:NOAC}

\begin{figure}[htbp]
  \centering
  \includegraphics[width =
    4in]{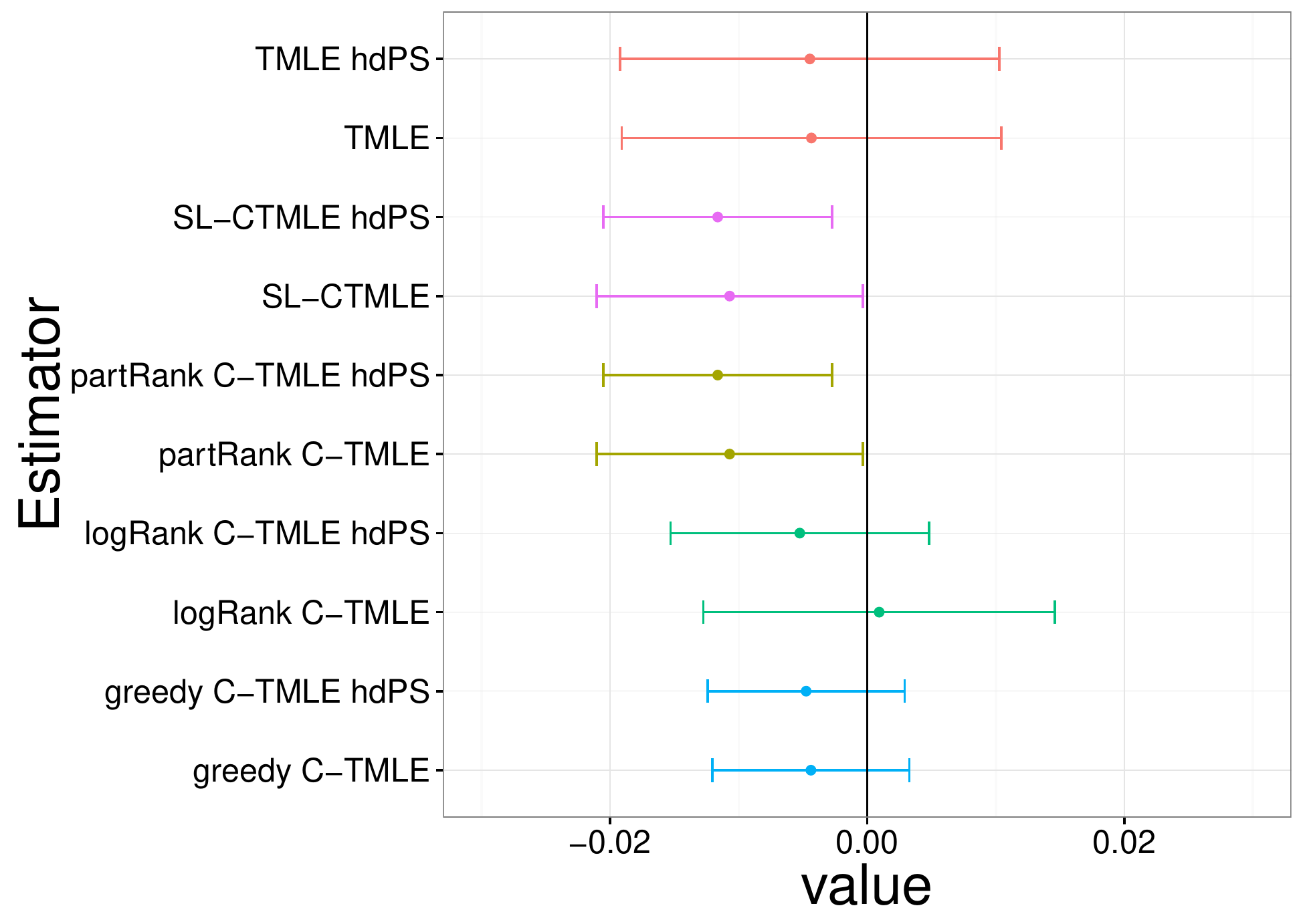}
    \caption{Point estimates  and 95\% CIs    yielded by
      the different TMLE and C-TMLEs built on the NOAC data set.}
  \label{fig:noac-ctmle}
\end{figure}

Figure~\ref{fig:noac-ctmle} shows  the point  estimates and 95\% CIs yielded by the different TMLE and C-TMLE estimators built on the
NOAC data  set. 

We observe more variability in the results than in  those presented in Appendix~\ref{subsec:NSAID}.



The  various  TMLE and  C-TMLEs   exhibit  similar results,  with  a
non-significant shift to the right for the latter. All the CIs contain zero.

\subsection{Results on the Vytorin Data Set}
\label{subsec:Vytorin}

\begin{figure}[htbp]
  \centering
  \includegraphics[width =
    4in]{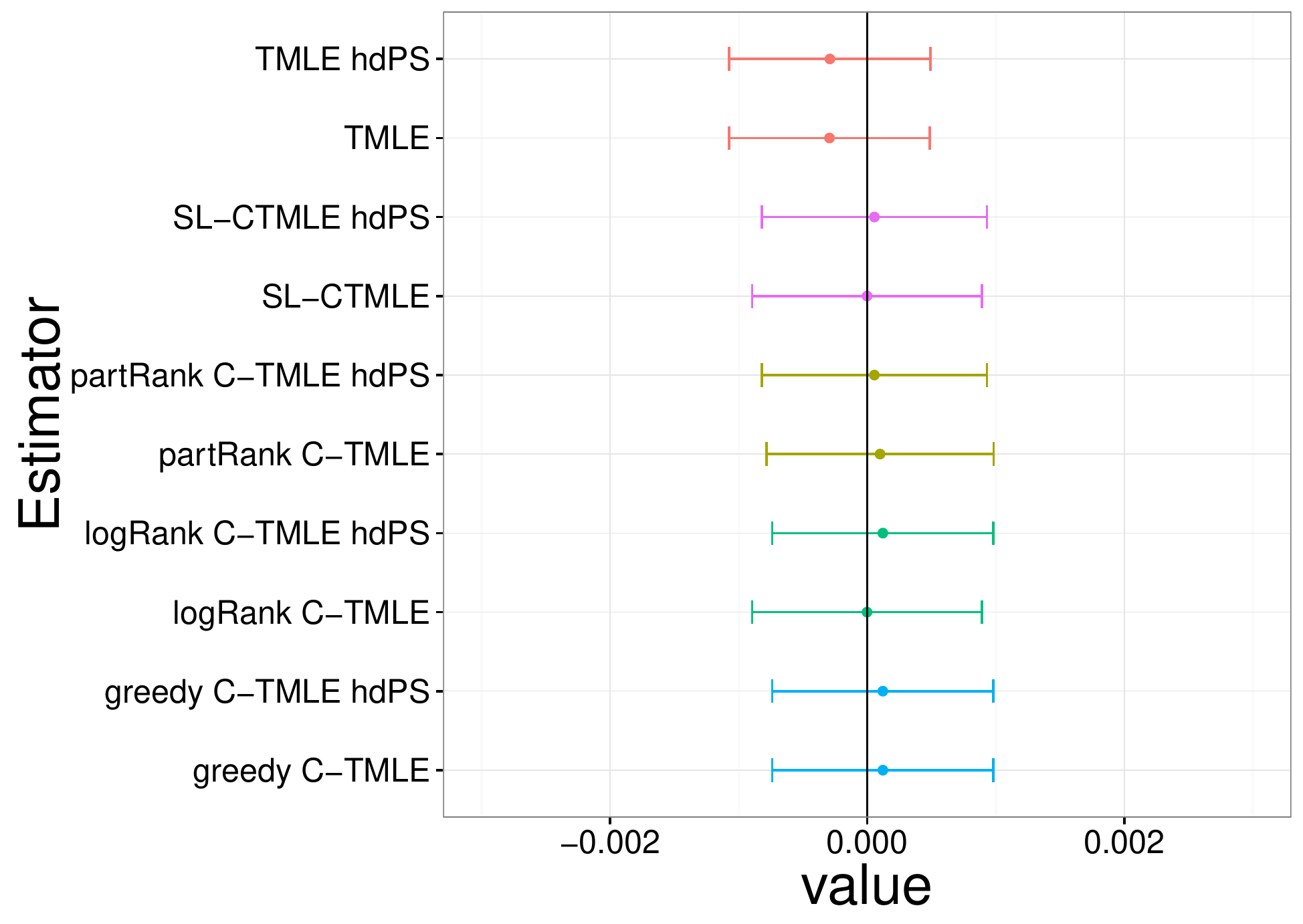}
    \caption{Point estimates  and 95\% CIs  yielded by
      the different TMLE and C-TMLEs built on the Vytorin data set.}
  \label{fig:vytorin-ctmle}
\end{figure}

Figure~\ref{fig:vytorin-ctmle} shows the point estimates and 95\% CIs  yielded by the different TMLE and C-TMLEs built on the
Vytorin data set.

The  various  TMLE and  C-TMLEs  exhibit  similar results,  with  a
non-significant shift to the right for the latter. All the CIs contain zero.

\end{document}